\documentclass[10pt,amsmath,amssymb,aps,nofootinbib,notitlepage,prd,superscriptaddress,twocolumn]{revtex4-1}

\usepackage[T1]{fontenc}
\usepackage{aas_macros,color,graphicx,hyperref,mathrsfs,orcidlink,subfigure}
\hypersetup{colorlinks=true,allcolors=blue}

\let\originalleft\left
\let\originalright\right
\renewcommand{\left}{\mathopen{}\mathclose\bgroup\originalleft}
\renewcommand{\right}{\aftergroup\egroup\originalright}

\newcommand{\ed}{\mathop{}\!\mathrm{d}}

\begin{document}

\title{Interferometric inference of black hole spin from photon ring size and brightness}

\author{Joseph R. Farah\,\orcidlink{0000-0003-4914-5625}}
\email{josephfarah@ucsb.edu}
\affiliation{Department of Physics, University of California Santa Barbara, Santa Barbara, California 93106, USA}
\affiliation{Las Cumbres Observatory, Goleta, California 93106, USA}

\author{Alexandru Lupsasca\,\orcidlink{0000-0002-1559-6965}}
\email{alexandru.v.lupsasca@vanderbilt.edu}
\affiliation{Department of Physics \& Astronomy, Vanderbilt University, Nashville, Tennessee 37212, USA}

\author{Eliot Quataert\,\orcidlink{0000-0001-9185-5044}}
\affiliation{Department of Astrophysical Sciences, Princeton University, Princeton, New Jersey 08544, USA}

\author{Michael D. Johnson\,\orcidlink{0000-0002-4120-3029}}
\affiliation{Center for Astrophysics — Harvard \& Smithsonian, 60 Garden Street, Cambridge, MA 02138, USA}
\affiliation{Black Hole Initiative at Harvard University, 20 Garden Street, Cambridge, MA 02138, USA}

\begin{abstract}
The $n=1$ photon ring is a full image of the astrophysical source around a black hole, produced by photons that execute $n\approx1$ half-orbit around the event horizon on their way to an observer.
The Black Hole Explorer (BHEX) is a proposed extension of the Event Horizon Telescope to space that will target the $n=1$ photon rings of the supermassive black holes M87${}^\ast$ and Sgr\,A${}^\ast$.
In this paper, we introduce a new interferometric observable that will be directly measurable on BHEX baselines and which admits a clear image-domain interpretation in terms of the photon ring brightness profile.
Across a wide range of semi-analytic equatorial emission models, we find that the azimuthal intensity profile of the ring can change depending on the astrophysics of the source, but its width $w_b$ is weakly sensitive to these details---much like the ring shape, which has previously been identified as a probe of the spacetime geometry.
Our survey suggests that interferometric measurements of the photon ring diameter and $w_b$ can place constraints (to $\lesssim\!20\%$) on the spin and inclination of a black hole with a known mass-to-distance ratio, such as Sgr\,A${}^\ast$.
State-of-the-art numerical simulations support this finding, paving the way to a precise photon-ring-based spin measurement for Sgr\,A${}^\ast$ with BHEX.
\end{abstract}

\maketitle

\section{Introduction}

The Event Horizon Telescope (EHT) has used an array of ground radio stations to observe the black holes M87${}^\ast$ and Sgr\,A${}^\ast$ via very-long-baseline interferometry (VLBI) \citep{M87I,SgrAI}.
The Black Hole Explorer (BHEX) is a mission under development to extend terrestrial (sub)millimeter VLBI to space by launching a satellite into Earth orbit at $\gtrsim20,000\,$km altitude \cite{BHEXMotivation,BHEXInstrument}.
With an effective aperture three times larger than the Earth, BHEX will take the sharpest images in the history of astronomy, enabling it to measure the first ($n=1$) photon rings around the supermassive black hole targets M87${}^\ast$ and Sgr\,A${}^\ast$ \cite{BHEXScience}, which so far remain unresolved \citep{Lockhart_2022,Tiede2022}.

The $n=1$ photon ring is a full image of the astrophysical source around a black hole, produced by photons that execute $n\approx1$ half-orbit around the horizon on their way to their observer \cite{Luminet1979,GrallaHolzWald,JohnsonLupsasca2020, GrallaLupsascaLensing}.
Since these photons probe the strong gravity of the spacetime just outside the horizon, the ring image that they produce carries information about the geometry of the black hole, including its mass, inclination and spin.
In particular, inferring the spins of the black holes M87${}^\ast$ and Sgr\,A${}^\ast$ from measurements of their $n=1$ photon rings is a primary goal for BHEX \cite{BHEXScience}.

The interferometric response to M87${}^\ast$ or Sgr\,A${}^\ast$ on the baselines sampled by an Earth-orbiting satellite is dominated by the $n=1$ ring.
This nearly circular ring has a typical diameter of $d\approx10M$ and a narrow width $w\lesssim M\ll d$, and its interferometric signature on baselines $1/d\ll u\ll1/w$ sampled by BHEX is essentially \emph{universal}: that is, the visibility is almost completely fixed by the ring size, shape, and brightness asymmetry, with scant dependence on its (unresolved) radial profile.
The ring produces a characteristic ringing signature whose periodicity and amplitude encode its angle-dependent diameter and brightness asymmetry, respectively.
The brightness asymmetry visibly varies with the rotation of the black hole and is clearly worth investigating as a tool for spin inference.

Here, we consider two visibility-domain observables that are directly measurable on the BHEX baselines where the $n=1$ subring dominates ($u\!\sim\!20$-$40\,\mathrm{G}\lambda$) and that together encode the geometry of the ring as well as its azimuthal brightness profile.
While this profile can rotate and change in ways that do depend on the astrophysics of the source, we find that its \emph{width} (illustrated in Fig.~\ref{fig:observable_construction}) is weakly sensitive to these details---much like the diameter itself, which has previously been identified as a probe of the spacetime geometry \cite{Cardenas2023}.

A na\"ive measure of the brightness asymmetry of the photon ring based on its visibility-amplitude envelope---e.g., $(V_{\max}+V_{\min})/(V_{\max}-V_{\min})$ as a function of baseline angle $\varphi$---is attractive but proves sensitive to astrophysical details and is difficult to estimate with low signal-to-noise ratio (SNR).
Instead, we integrate the squared visibilities over the $n=1$-dominated annulus in the $(u,v)$ plane to obtain an angular profile $b_\varphi\!\equiv\!\int_{u_-}^{u_+}\!\!|V(u,\varphi)|^2\,\ed u$, whose full width at half maximum we denote by $w_b$ (see Fig.~\ref{fig:observable_construction}).
Physically, $b_\varphi$ tracks the azimuthal distribution of subring brightness, and $w_b$ provides a robust, directly interferometric proxy for the width of that distribution.

We pair $w_b$ with a geometric observable $d_+$, the maximum of the angle-dependent diameter $d_\varphi$ of the $n=1$ subring, inferred from the periodicity of the Bessel-like ringing in the visibility amplitude---see Eq.~\eqref{eq:visamp_approximation}.
Across semi-analytic equatorial emission models, we find that $w_b$ is primarily controlled by inclination and is only weakly affected by nuisance emission parameters, whereas $d_+$ varies strongly with spin at low inclinations and exhibits different (and largely orthogonal) degeneracies.
A joint measurement of $(d_+,w_b)$ therefore constrains spin and inclination simultaneously.

This spin-inference method is typically subject to uncertainties at the $\lesssim\!20\%$ level for sources with known mass-to-distance ratio such as Sgr\,A${}^\ast$, consistent with our survey and with tests against state-of-the-art simulations using general-relativistic magnetohydrodynamics (GRMHD; \cite{Gammie_2003}).
A major advantage of this technique is that it can be carried out entirely within the visibility domain (without requiring reconstruction of an image from its sparsely sampled Fourier transform) while still relying on data with a clear geometric interpretation.

Our approach complements and extends prior image-domain and geometric strategies.
Image-fitting and ``photogrammetry'' frameworks can recover spin in controlled settings \cite{BlackHolePhotogrammetry}, and equatorial semi-analytic models such as \texttt{KerrBAM} reproduce many MAD features (though not SANE) and constrain inclination but find spin difficult to pin down \cite{KerrBAM}.
Measuring multiple subring diameters can in principle break shape degeneracies and yield $\sim$10\% spin constraints, albeit with strong mass degeneracy \cite{Broderick2022}.
Machine-learning analyses of photon-ring images also indicate $\lesssim$10\% spin precision in favorable cases \cite{Farah:PR1}.
Additional channels---e.g., linear polarization of the ring to break mass-spin degeneracies at low inclination \cite{LinearPolarization} and tests based on detailed ring-shape systematics \cite{Paugnat2022}---provide valuable cross-checks.
Most closely related to our goals, recent geometric attempts using $n=1$ diameter and asymmetry alone reported limited constraining power in the BHEX regime, especially at moderate spins \cite{Keeble2025}; by contrast, our visibility–integrated $w_b$ is both easier to measure at low SNR and more robust to astrophysical variations, particularly for $\theta_{\rm o}\lesssim50^\circ$.

Finally, we outline assumptions and practicalities.
Our semi-analytic survey and GRMHD validation focus on primarily equatorial emission (appropriate to \texttt{KerrBAM} and many MAD simulations) and on baselines where the $n=1$ subring dominates; off-equatorial or highly turbulent emission can broaden the $(u,v)$ support and introduce biases that we quantify below.
For Sgr\,A${}^\ast$, where $M/D$ is known, $(d_+,w_b)$ alone can yield meaningful spin constraints.
For M87${}^\ast$, combining $w_b$ with a mass-sensitive observable---such as a diameter ratio (e.g., $d_1/d_0$) or independent polarization constraints---can break residual degeneracies and potentially enable joint mass-spin inference with high precision.

In this letter, we investigate two interferometric observables of the $n=1$ subring which, when measured jointly, can constrain spin to $\lesssim20\%$ or better at low inclinations.
We begin in \autoref{sec:response} by briefly reviewing the response of an interferometer to a thin ring on the sky.
In \autoref{sec:construction}, we motivate and construct $d_+$ and $w_b$ (a measure of maximum diameter and of brightness-profile width) and chart their behavior across spins and inclinations.
In \autoref{sec:GRMHD_validate}, we validate their performance on a suite of GRMHD simulations.
We conclude with limitations and prospects in \autoref{sec:conclusions}.

\section{Interferometric signature of a thin ring}
\label{sec:response}

Here, we review the interferometric response to a thin ring.
Following Gralla \cite{Gralla2020}, we do not assume that the ring is circular, but rather allow its shape to describe any closed convex curve.
Thus, we consider an arbitrary convex ring of angle-dependent diameter $d_\varphi\sim d$ and width $w\ll d_\varphi$, with local transversely-integrated intensity $\mathcal{I}$.
Then, on baselines $u$ satisfying $1/d\ll u\ll 1/w$,  the visibility amplitude $|V|$ can be approximated as \cite{Gralla2020}
\begin{align}
    \label{eq:visamp_approximation}
    |V|\approx\frac{1}{\sqrt{u}}\sqrt{\left(\alpha_{\varphi}^{\rm L}\right)^2+\left(\alpha_{\varphi}^{\rm R}\right)^2+2\alpha_{\varphi}^{\rm L}\alpha_{\varphi}^{\rm R}\sin\left(2\pi d_{\varphi}u\right)},
\end{align}
where the angle-dependent functions
\begin{align}
    \alpha_{\varphi}^{({\rm L/R})}=\left.\mathcal{I}\sqrt{\mathcal{R}}\right|_{\rm L/R}
\end{align}
encode the integrated intensity $\mathcal{I}$ and radius of curvature $\mathcal{R}$ around the ring.
A primary feature of this interferometric signature is the periodicity $1/d_\varphi$ of the sinusoid, which can be measured and used to infer the diameter of the ring.
Additionally, the visibility amplitude oscillates within an envelope given by 
\begin{align}
    V_{\max}=\frac{\alpha_{\varphi}^{\rm L}+\alpha_{\varphi}^{\rm R}}{\sqrt{u}},\quad
    V_{\min}=\frac{\left|\alpha_{\varphi}^{\rm L}-\alpha_{\varphi}^{\rm R}\right|}{\sqrt{u}}.
\end{align}
Thus, measurements of the damped oscillation can also yield information about the source integrated intensity $\mathcal{I}$, particularly for rings of near-constant curvature $\mathcal{R}$.

\section{Construction of interferometric observables
}
\label{sec:construction}

Recent works \cite{Gralla:PR1,Gralla:PR2,Farah:PR1} have shown that the azimuthal profile and maximum diameter of the $n=1$ subring correlate with spin in both the image and visibility domains. Forming images with limited coverage can be challenging, so here we construct and investigate Fourier analogs of these spin-distinguishing geometric observables.

\begin{figure*}
    \centering
    \includegraphics[width=\linewidth]{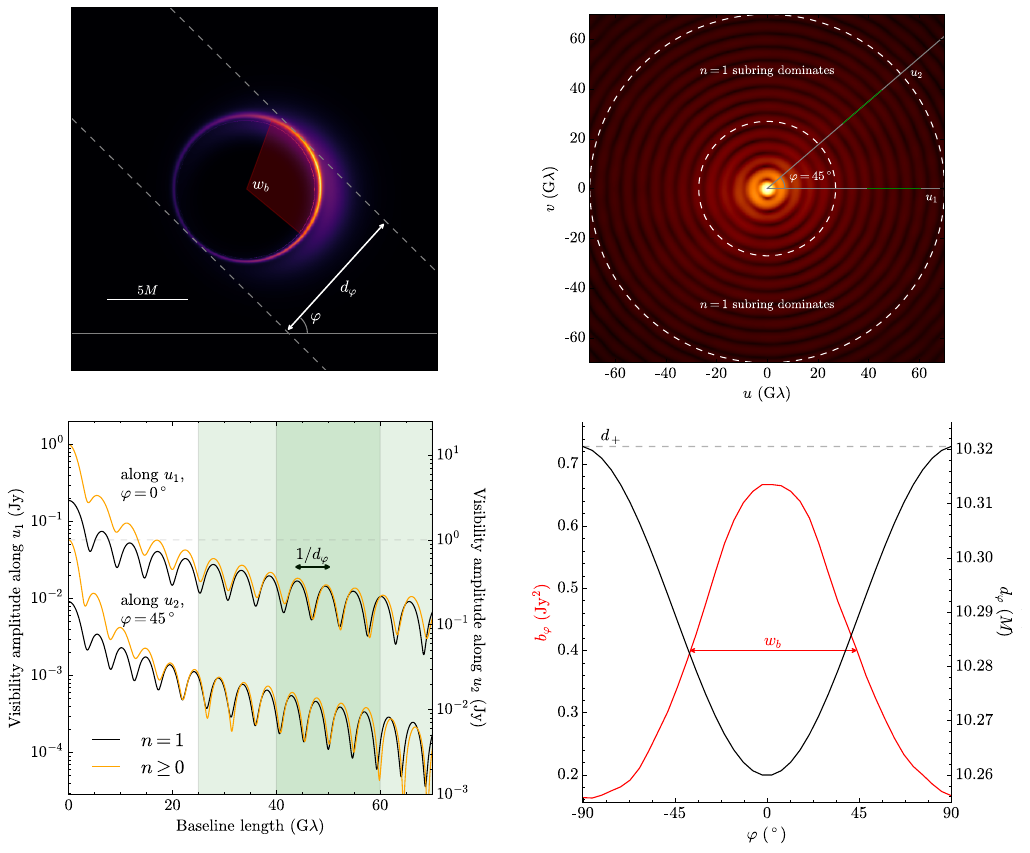}
    \caption{Demonstration of how the region of dominance of the $n=1$ subring arises in the Fourier transform of an image containing multiple subrings. \textit{(top left)} An image containing a simulated black hole. In this image, at least 3 subrings are clearly visible: (i) the $n=0$ subring, visible as broad diffuse flux particularly on the right side of the image; (ii) the $n=1$ subring of size $d_\varphi$ at angle $\varphi$, a sharp ring of light and the most dominant image feature, and (iii) the $n=2$ subring, concentrically nested within the first subring and visible as a much thinner and dimmer ring of light. \textit{(top right)} the absolute value of the Fourier transform of the image in the first panel. Using properties derived in \cite{JohnsonLupsasca2020}, we expect the $n=1$ subring to dominate the Fourier transform in the region bounded by the dashed white lines. The gray rays correspond to $u_1$ and $u_2$, which will be examined in the following panel. \textit{(bottom left)} Cross-sections of the Fourier transform along the rays $u_1$ and $u_2$, showing the characteristic Bessel function behavior. If we compute the $n=1$ subring visibility function and compare it to the full Fourier transform, we see it dominates in the light green region, with the most sigificant domination (i.e., almost all of the emission is generated by the $n=1$ photon ring) occurring in the darker green region. The spacing of the peaks in this region scales as $1/d_\varphi$ and therefore is a direct measurement of $d_\varphi$ of the $n=1$ subring. We can also integrate the visibilities in this region to compute $b_\varphi$ as described in \autoref{sec:construction}. \textit{(bottom right)} construction of the diameter $d_\varphi$ and $b_\varphi$ for all distinct angles in the Fourier plane, computed over the dark green shaded region in the bottom left panel. Both functions look Gaussian- or sinusoid-like. We compute the maximum diameter $d_+$ (dashed line) and the full-width half-maximum (FWHM) of $b_\varphi$ as our interferometric observables.}
    \label{fig:observable_construction}
\end{figure*}

\cite{JohnsonLupsasca2020} showed that emission from $n=1$ subring with size $d$ and width $\alpha$ is expected to dominate the Fourier transform over a finite range of baseline lengths $1/d \ll u \ll 1/\alpha$ probeable by BHEX for M87${}^\ast$ and Sgr A*. We can exploit this phenomenon to make inferences about the $n=1$ subring while ignoring complications from surrounding astrophysical phenomena by constructing our observables in this $(u, v)$ region of dominance. A demonstration of the behavior of the subring Fourier transform in this region of dominance is shown in \autoref{fig:observable_construction}.

We begin by constructing the observable of maximum diameter, $d_+$. Following the treatment of \cite{Gralla2020} and \cite{JohnsonLupsasca2020}, the diameter of the $n=1$ subring in a particular direction can be inferred by measuring the frequency of the visibility amplitude oscillation (see \autoref{eq:visamp_approximation}) over the region of dominance. We estimate the diameter $d_\varphi$ in every direction and construct our first interferometric observable by taking the maximum inferred diameter $d_+$. 

We next examine the azimuthal brightness profile. \cite{Farah:PR1} heuristically found that variation of this brightness profile tracked the inclination of the black hole, particularly at low spins and inclinations. This finding can be intuitively motivated by considering that the dominant contribution to variation in the angular brightness profile is the redshift of observed photons due to the velocity of the orbiting fluid. Beginning with equation B13 from \cite{AART}, we express the redshift of a photon emitted from a circularly orbiting fluid element as, 
\begin{align}
    g = \frac{E}{-p_\mu u^\mu} = \frac{1}{u^t(1-\lambda\Omega)},
\end{align}
where $\lambda = -p_\phi/p_t$ is the specific angular momentum, and $\Omega = u^\phi/u^t$ is the angular velocity of the circular-equatorial emitter. For an observer at inclination $\theta_{\rm o}$, $\lambda = -\alpha\sin\theta_{\rm o}$, where $\alpha$ is the horizontal image coordinate. Passing to polar image coordinates $(\rho, \varphi)$, $\lambda = -\rho\cos\varphi \sin\theta_{\rm o} = -r_s\cos\varphi \sin\theta_{\rm o}$, where in the last step we use the fact that $r_s = \rho$ in flat space. Moreover, for a Keplerian orbit, $v = r_s\Omega$, and $u = u^t(\partial_t+\Omega\,\partial_\phi)$, where $u^t=(1-v^2\sin^2\theta_{\rm o})^{-1/2}$. Therefore, for a ring of radius $r_s$ in flat space, with on-sky azimuthal angle $\varphi$, the redshift becomes
\begin{align}
    g = \frac{\sqrt{1-v^2\sin^2\theta_{\rm o}}}{1+v\cos\varphi\sin\theta_{\rm o}}.
    \label{eq:redshift_final}
\end{align}
At low to moderate spins and most inclinations, \autoref{eq:redshift_final} tracks the azimuthal brightness profile well. In particular, when normalized to remove overall offsets or scalings in intensity, the width (i.e., full width half maximum) of the azimuthal brightness profile scales directly with inclination, motivating its use as an observable. 

A natural measure of brightness asymmetry $\mathcal{A}$ in Fourier space is, following \cite{Gralla2020},
\begin{align}
    \mathcal{A} = \frac{V_{\text{max}} + V_{\text{min}}}{V_{\text{max}} - V_{\text{min}}}.
\end{align}
However, in testing on simple semi-analytic toy models, we find that in the range of BHEX space baselines ($\approx 20\text{-}35 \ \text{G}\lambda$) this quantity varies more with the astrophysics of the source than with the properties of the black hole spacetime. 

Therefore, we propose an alternative analogous observable: the integrated sum of the squared visibility amplitudes on the region of $n=1$ subring dominance. Consider the quantity $B_\varphi^{(k)}$ which tracks (as a function of angle $\varphi$) moments of the total intensity of a projection $\mathcal{P}_\varphi$ of an image $I(s)$ of the $n=1$ photon ring,
\begin{align}
    B_\varphi^{(k)} = \int_{-\infty}^{\infty} ds \ \big[\mathcal{P}_{\varphi}I(s)\big]^k.
\end{align}
Using Plancherel's theorem, we can relate the interferometer response to the photon ring to the $k=2$ moment of $B_\varphi^{(k)}$, i.e., 
\begin{equation}
    \int_{-\infty}^{\infty} du \ |V(u, \varphi)|^2 = \int_{-\infty}^{\infty} dr \ |\mathcal{P}_\varphi I(r, \varphi)|^2 = B_\varphi^{(2)},
    \label{eq:bvarphi_vis}
\end{equation} 
where $V(u, \varphi)$ is the visibility of the source measured by an interferometer. However, in order to directly probe the $n=1$ subring, we may only measure on a range of frequencies $u_- < u < u_+$ where the $n=1$ subring dominates. We therefore define a related new quantity $b_\varphi$ which only probes this range of baselines,
\begin{align}
    b_\varphi \equiv \int_{u_-}^{u_+} du \ |V(u, \varphi)|^2.
\end{align}
We next assume we are probing baseline lengths far from the origin, and use the approximation derived in Eq. (5) of \cite{Gralla2020}, reproduced in \autoref{eq:visamp_approximation}. Under these conditions, 
\begin{equation}
    b_\varphi \propto (\alpha_\varphi^L)^2 + (\alpha_\varphi^R)^2 \propto (I_\varphi^{L})^2 + (I_\varphi^{R})^2,
\end{equation}
where in the last step we have exploited the fact that the $n=1$ subring has an approximately constant radius of curvature $\mathcal{R}$. The quantity $b_\varphi$ can be measured relatively easily, and has an approximately Gaussian-like profile. We characterize this profile with our second interferometric observable $w_b$ equal to the full-width half-maximum (FWHM) of $b_\varphi$.  

We next seek to characterize the behavior of $d_+$ and $w_b$ for the $n=1$ subring at a range of spins and inclinations. Additionally, we expect that characteristics of the azimuthal brightness distribution will vary with properties of the accretion and fluid flow, which may complicate spin inference. Following \cite{Farah:PR1}, we use \texttt{KerrBAM} \citep{KerrBAM} to construct a simulation grid of black holes at combinations of 50 spins ($-0.99 < a < 0.99$) and inclinations ($0 < \theta_{\mathrm{o}}^\circ < 75^\circ$). Additionally, for each combination of spin and inclination, we vary: $0.05 < \beta < 0.95$, the fluid velocity in the zero angular momentum observer (ZAMO) frame; $\pi/5 < \iota < \pi/2$, the vertical angle of the magnetic field vector; $-\pi/2 < \eta < \pi/2$, the equatorial angle of the magnetic field vector; and $-\pi/2 > \chi > -3\pi/2$, the equatorial fluid velocity angle. At each spin magnitude and inclination, we calculate values of $d_+$ and $w_b$ over the range $25 \mathrm{\ G\lambda} < u < 35 \mathrm{\ G\lambda}$ for each combination of the fluid parameters, forming a distribution of values from which we compute a discrete median and standard deviation. We expand on the details of observable measurement in \autoref{appendix:observable_measurement}. 

\begin{figure*}
    \centering
    \includegraphics[width=0.425\linewidth]{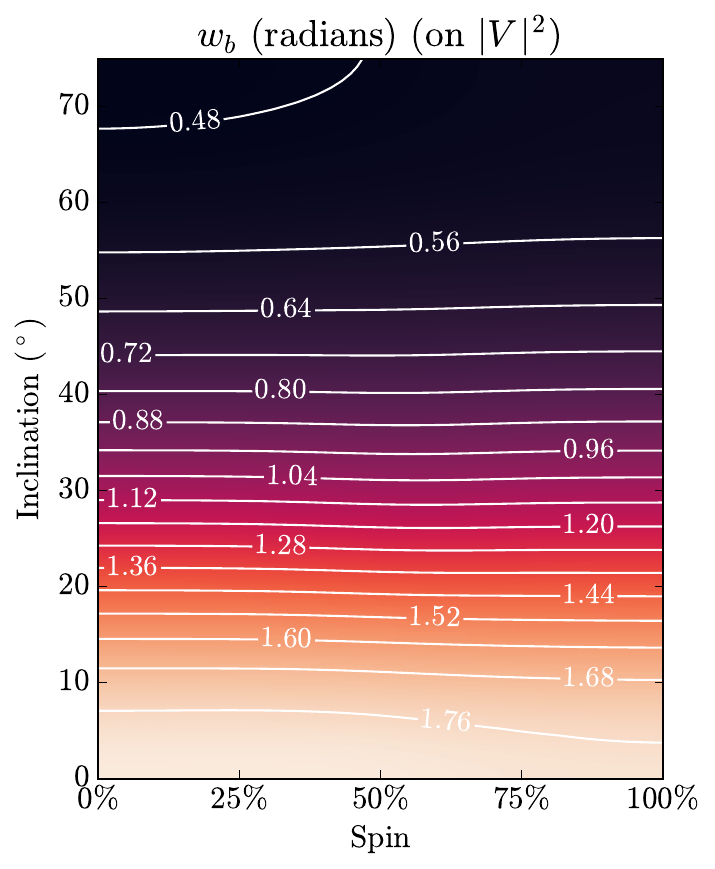}
    \includegraphics[width=0.425\linewidth]{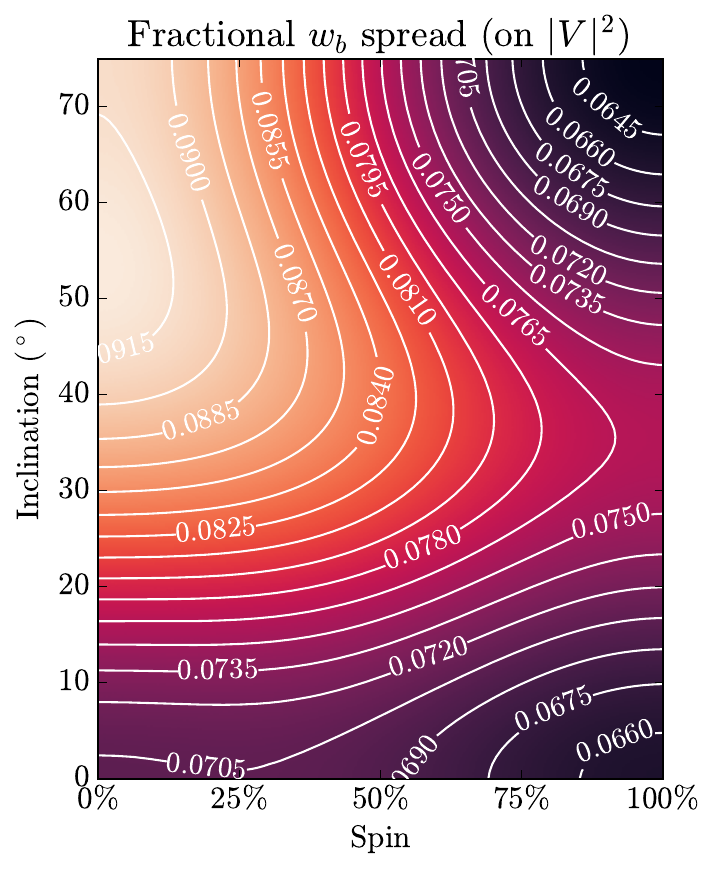}
    \caption{Median (\textit{left}) and spread (\textit{right}) of values of $w_b$ computed from the simulation grid described in \autoref{sec:construction}. By eye, $w_b$ tends to vary strongly with inclination for the entire Kerr parameter space. The spread in values generated by varying properties of the fluid flow is generally $\approx8\%$ of the value of $w_b$ at inclinations $\theta_{\rm o} < 50^\circ$.}
    \label{fig:omega_b}
\end{figure*}

\begin{figure*}
    \centering
    \includegraphics[width=0.45\linewidth]{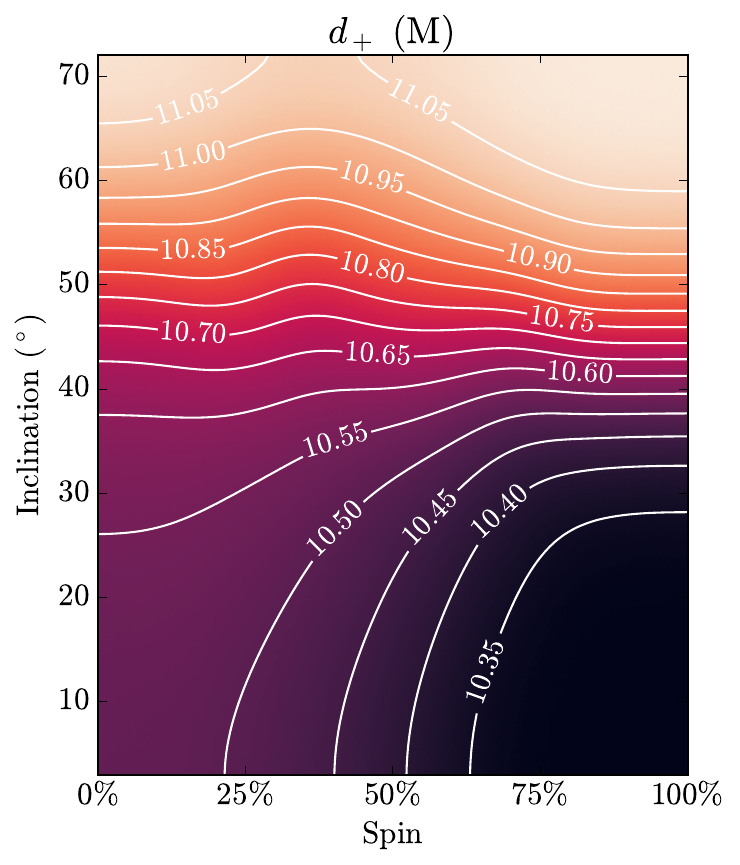}
    \includegraphics[width=0.45\linewidth]{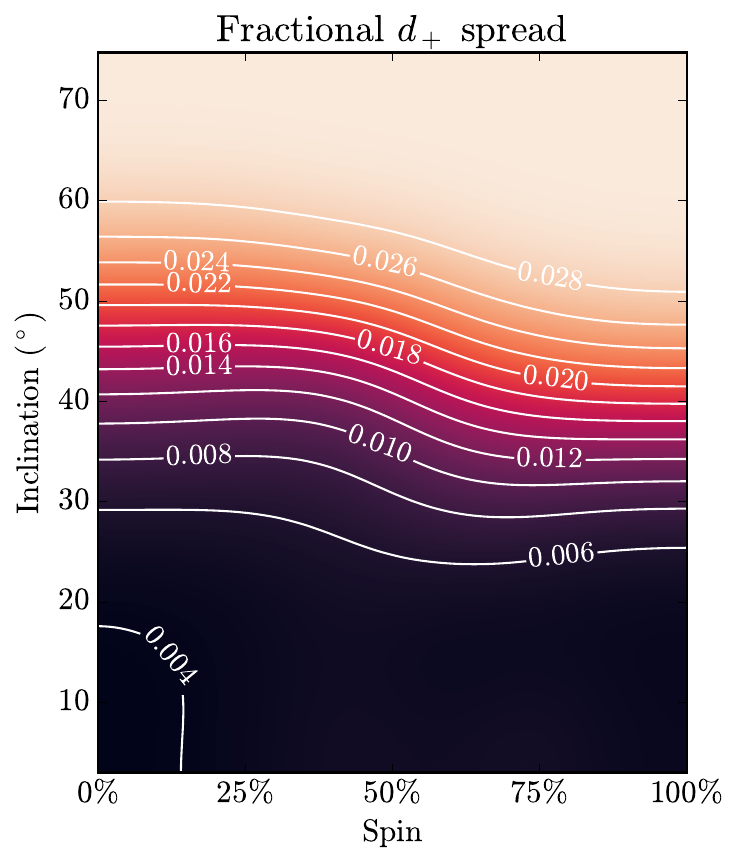}
    \caption{Median (\textit{left}) and spread (\textit{right}) of values of $d_+$ computed from the simulation grid described in \autoref{sec:construction}. By eye, $d_+$ tends to vary strongly with spin at inclinations $\lesssim 40^\circ$. The spread in values generated by varying properties of the fluid flow at these inclinations is generally $\lesssim 1\%$ of the value of $d_+$.}
    \label{fig:dmax}
\end{figure*}

The resulting distributions of $w_b$ and $d_+$ are shown in \autoref{fig:omega_b} and \autoref{fig:dmax}, respectively. Across all tested spins and inclinations, we find that $w_b$ trends strongly with inclination, and the modification of the parameter due to non-GR astrophysical phenomena (i.e., $\beta$ and $\iota$) is less than a $10\%$ effect on average. By contrast, $d_+$ is largely orthogonal to $w_b$ at inclinations $\theta_{\mathrm{o}} \lesssim 50^\circ$, and varies strongly with spin. Above inclinations of $\sim50^\circ$, $d_+$ tracks almost directly with inclination. In the region of the Kerr space where $d_+$ maps to spin, we find variation significantly less than $1\%$ with changes to $\beta$ and $\iota$. 

We also investigate the impact of additional parameters and choices on the resulting distributions of $d_+$ and $w_b$. First, we note that $d_+$ and $w_b$ display significant symmetry in spin (i.e., the values do not vary significantly with the sign of the spin). The variation with sign is a $\lesssim1\%$ effect, which we incorporate into the fractional spread (\autoref{fig:dmax}), and consider only the magnitude of the spin $|a|$. Second, we remark that more parameters than the ones varied here ($a, \theta_{\rm o}, \beta, \eta, \iota,$ and $\chi$) are expected to have an impact on the $d_+$ and $w_b$ quantities. For example, we find that varying the characteristic radius of the emission profile ($R$ in \cite{KerrBAM}) can significantly modify the measured value of $d_+$. Additionally, the synchrotron cross product index ($\alpha_\xi$ in \cite{KerrBAM}) and spectral index ($\alpha_\nu$) moderately modify the azimuthal brightness profile beyond what is probed in our simulation library. Finally, our analysis of $d_+$ and $w_b$ is additionally limited in that we assume equatorial emission (by nature of the \texttt{KerrBAM} toy model) and dominance of the $n=1$ subring.

\section{Validation on GRMHD simulations}
\label{sec:GRMHD_validate}

We next validate the effectiveness of our spin characterization method using general relativistic magnetohydrodynamic (GRMHD) simulations. Semi-analytic engines such as \texttt{KerrBAM} are powerful tools for studying the subring behavior in idealized conditions, but the underlying toy models are too simple to form the basis for a meaningful procedure validation. By contrast, GRMHD simulations are a best-guess at what a real horizon-scale environment may look like. Full numerical GRMHD simulations are far more complex and realistic than semi-analytic simulation engines, and incorporate many features likely to cause complications in a real observation (e.g., time variability). Performing inference on time-variable data is still an open challenge with more required to solve. However, these features can be time-averaged over long timescales to reduce their presence in the image. In this way, testing on GRMHD simulations can provide an accurate estimation of how a spin measurement procedure may perform on a real observation. 

To construct a realistic test suite, we generate GRMHD simulations at inclinations of $30^\circ$ and $10^\circ$, typically a challenging range for performing spin estimations, and aligned with inclination estimates for M87${}^\ast$ ($10\textrm{-}20^\circ$). In the construction of our test suite, we varied (in addition to the spin of the black hole): $1 \leq R_{\mathrm{high}} \leq 160$, the electron heating parameter in the weakly magnetized disk; $1 \leq R_{\mathrm{low}} \leq 10$, the electron heating parameter in the highly magnetized jet; and the poloidal magnetic field strength, varied between a fully magnetically-arrested disk (MAD) or standard and normal evolution (SANE). The GRMHD simulations were generated for spins with $|a|=0.25, 0.50$ and $0.93$. The simulations are then time-averaged over $\sim1.5\times10^4$ gravitational timescales $t_g\equiv GM/c^3$ (for M87* this is $t_g\sim9$ hours, for Sgr A* this is $t_g\sim20$ seconds). 

We perform measurements of $w_b$ and $d_+$ in a similar fashion to the application on the KerrBAM simulation grid, with some minor modifications due to the increased complexity of the GRMHD (\autoref{appendix:observable_measurement}). We time-average each GRMHD simulation and obtain a single representative image of the black hole and the surrounding environment. We then Fourier transform this image and construct $w_b$ and $d_+$ in the region of dominance of the $n=1$ subring, based on the properties of the simulation (demonstrated in \autoref{fig:grmhd_demo}, elaborated on in \autoref{appendix:observable_measurement}). A joint posterior distribution is constructed for $a$ and $\theta_{\rm o}$ by computing the likelihood of each $(a, \theta_{\rm o})$ combination based on the spread (standard deviation per $(a, \theta_{\rm o})$ bin) predicted by \texttt{KerrBAM}. The median of the marginal posterior distribution for each parameter is taken to be the prediction for each test.  

\begin{figure*}
    \centering
    \includegraphics[width=0.9\linewidth]{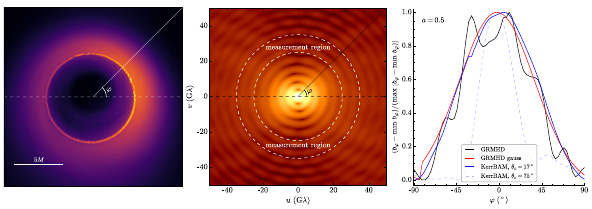}
    \caption{Demonstration of the extraction of $w_b$ as performed on GRMHD simulations. (\textit{left}) a time-averaged GRMHD simulation of a black hole with $a=0.5$ and $\theta_{\rm o} = 17^\circ$. (\textit{center}) Visibility amplitudes of the same simulation. The BHEX instrument is expected to achieve a maximum baseline length of $\sim35 \text{G}\lambda$, so we perform our measurement from $25 \text{G}\lambda$ (where the $n=1$ subring is expected to begin dominating) out to the maximum BHEX baseline length. (\textit{right}) We compute the $b_\varphi$ quantity for the simulated image (solid black line) following the procedure in \autoref{sec:construction}. The function is well-approximated by a Gaussian (solid red line). When compared to the $b_\varphi$ quantity computed on the corresponding \texttt{KerrBAM} image, the spreads (analogous to $w_b$) are consistent. Importantly, the spread in $b_\varphi$ at $17^\circ$ is also clearly distinct from the spread when computed on a much higher inclination \texttt{KerrBAM} image (light blue dashed line), making the $w_b$ quantity an effective proxy of inclination.  }
    \label{fig:grmhd_demo}
\end{figure*}

\begin{figure*}
    \centering
    \includegraphics[width=0.31\linewidth]{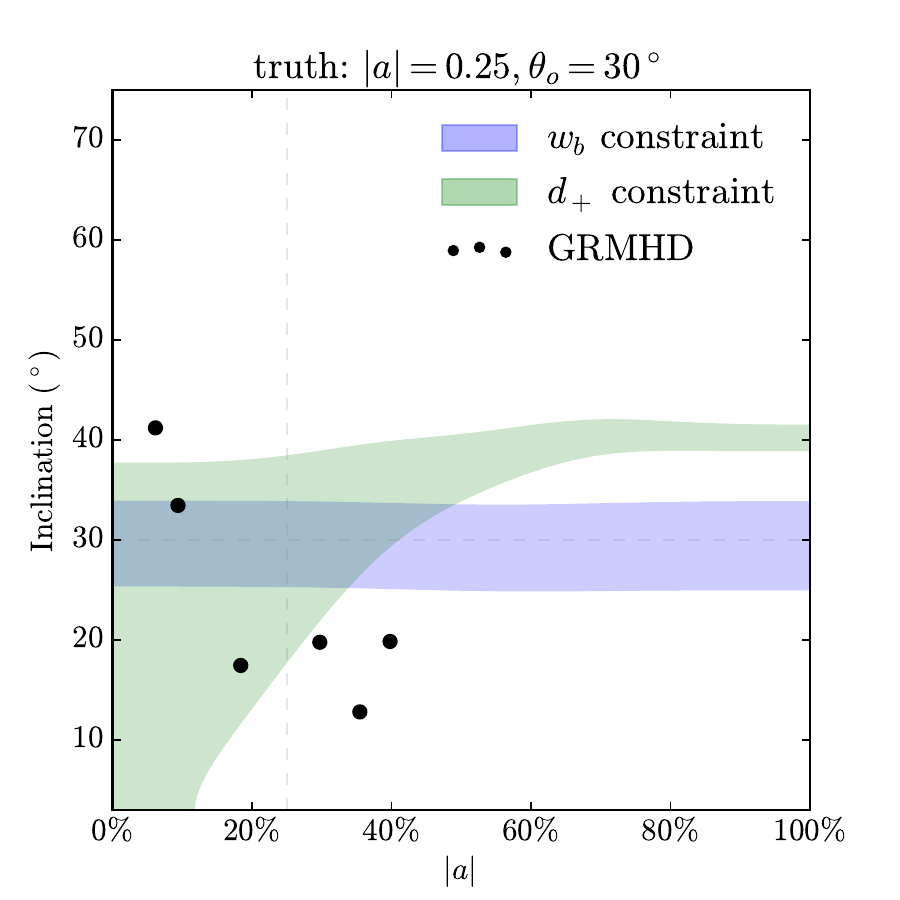}
    \includegraphics[width=0.31\linewidth]{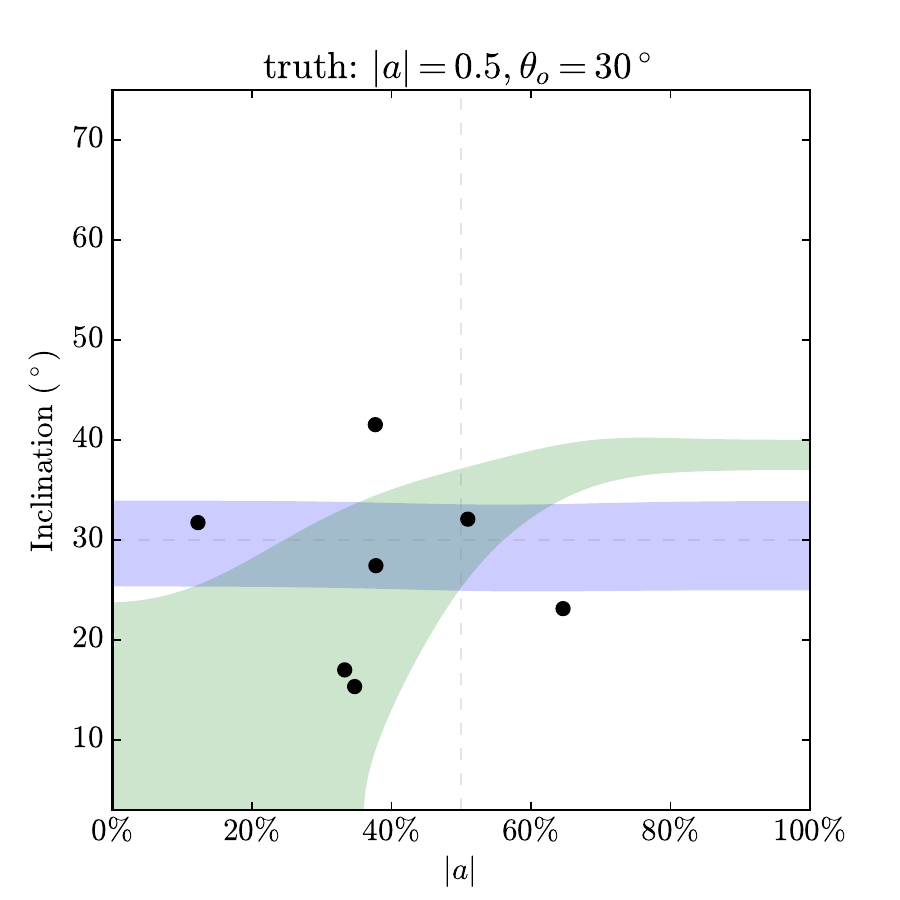}
    \includegraphics[width=0.31\linewidth]{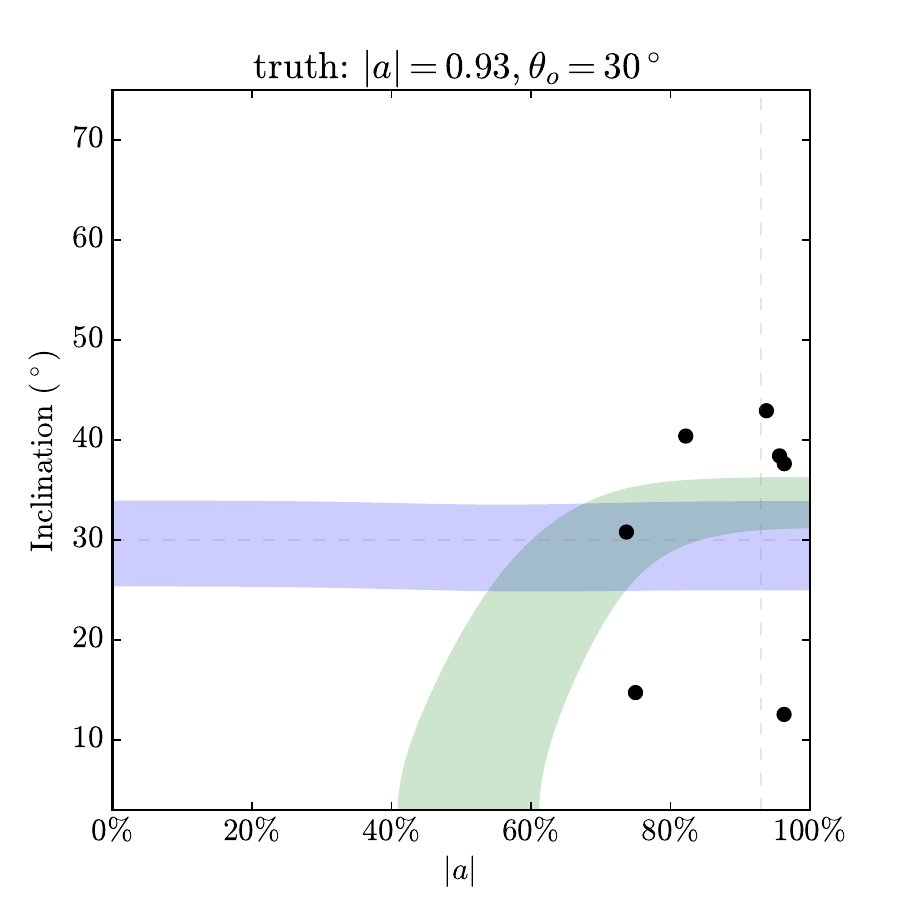}
    \caption{Ability of $d_+$ and $w_b$ to constrain spin and inclination when applied to GRMHD simulations, for $\theta_{\rm o}=30^\circ$. The bands correspond to average value and spread of the $w_b$ and $d_+$ parameters as computed from the \texttt{KerrBAM} grid. Black dots represent an estimate of spin and inclination based on measurements of $d_+$ and $w_b$ on the GRMHD simulation. While not centered on the truth values (dashed gray lines), the extractions on GRMHD simulations are consistent with the semi-analytic \texttt{KerrBAM} distributions to $\approx1\sigma$. } 
    \label{fig:inc30_grmhd}
\end{figure*}

\begin{figure*}
    \centering
    \includegraphics[width=0.31\linewidth]{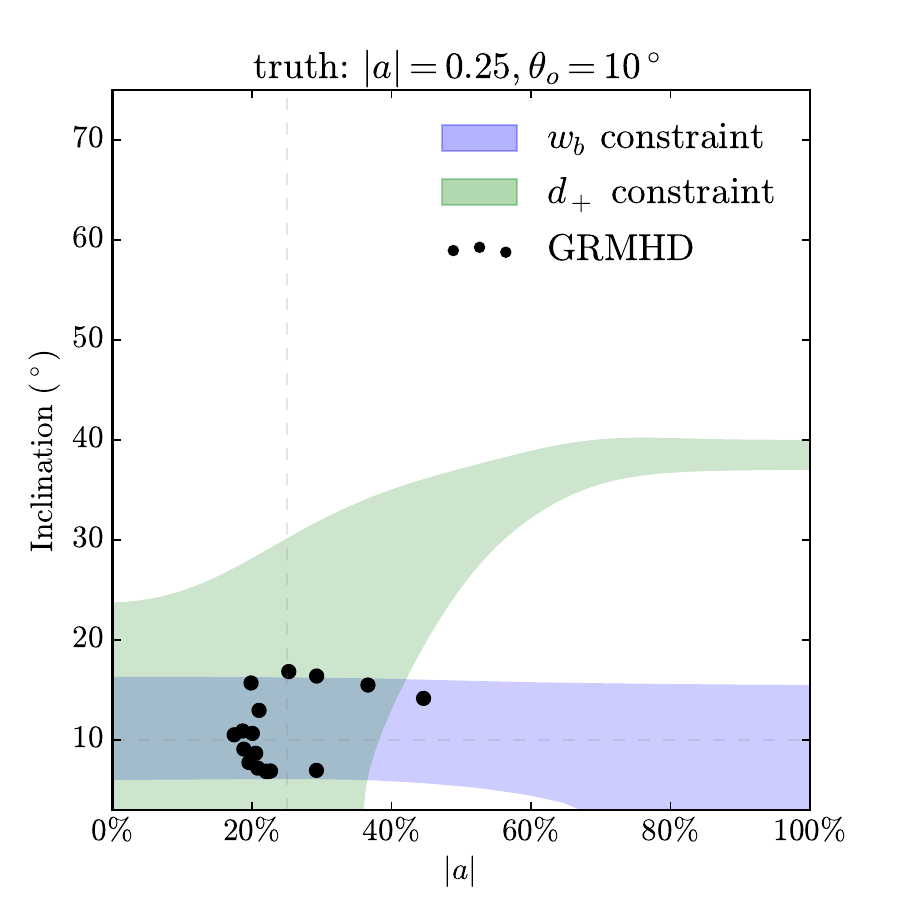}
    \includegraphics[width=0.31\linewidth]{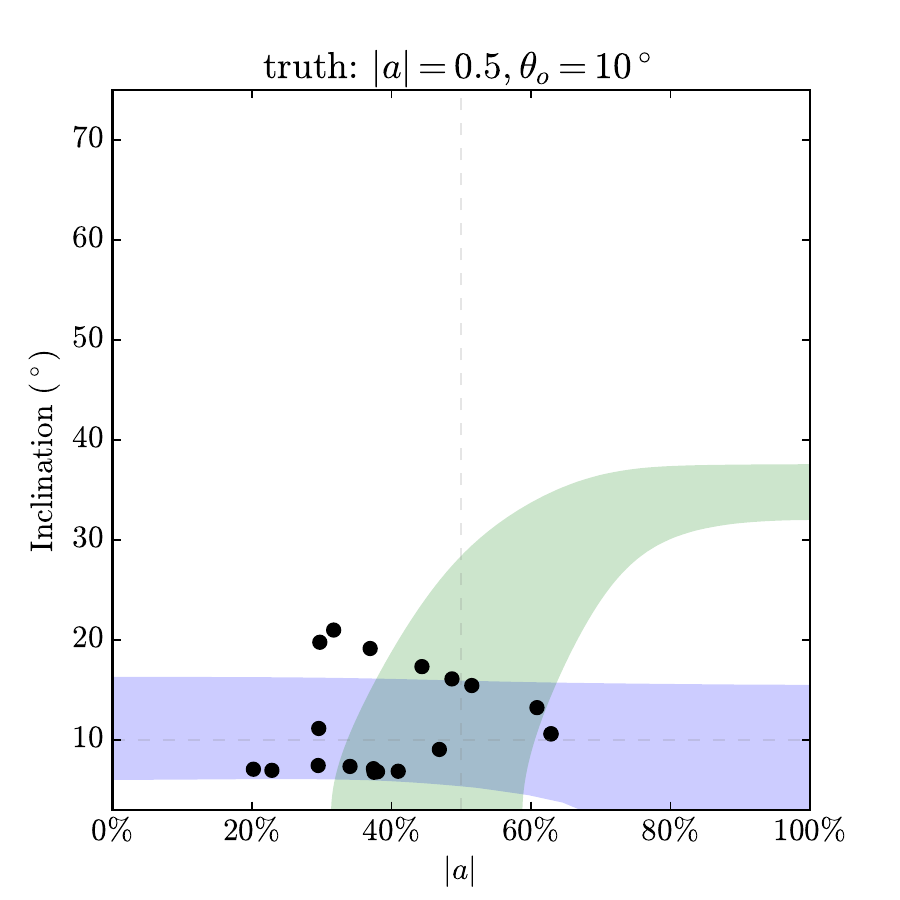}
    \includegraphics[width=0.31\linewidth]{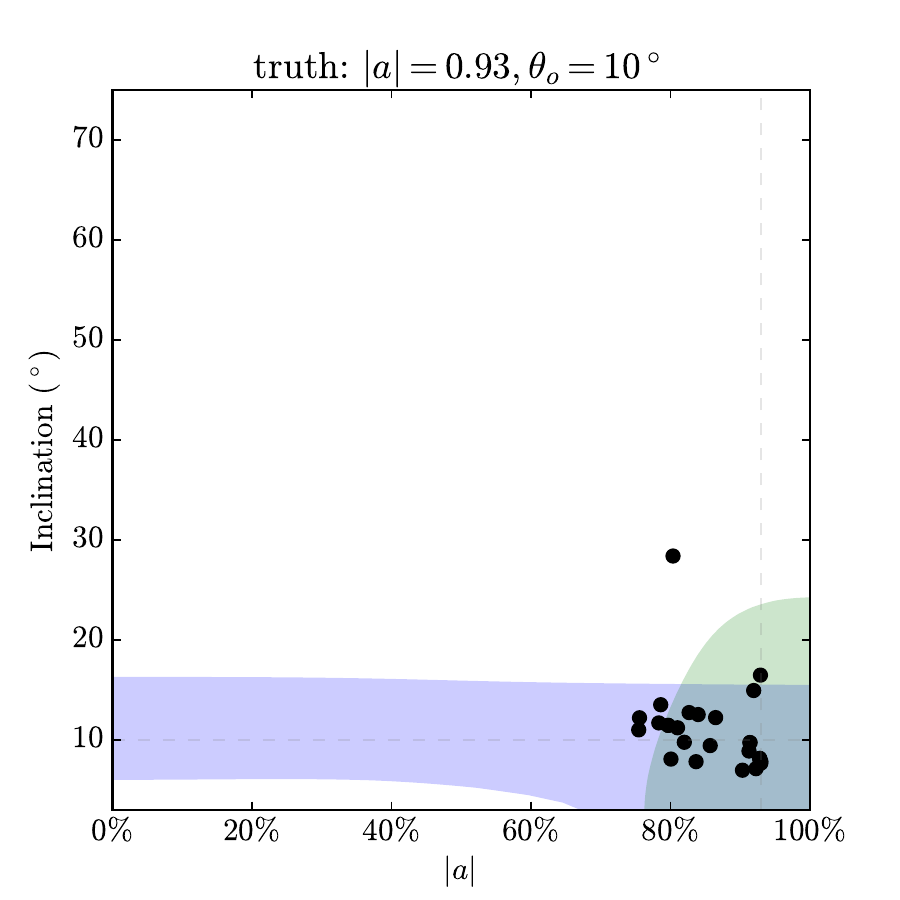}
    \caption{Ability of $d_+$ and $w_b$ to constrain spin and inclination when applied to GRMHD simulations, for $\theta_{\rm o}=10^\circ$. The bands correspond to average value and spread of the $w_b$ and $d_+$ parameters as computed from the \texttt{KerrBAM} grid. Black dots represent an estimate of spin and inclination based on measurements of $d_+$ and $w_b$ on the GRMHD simulation. While not centered on the truth values (dashed gray lines), the extractions on GRMHD simulations are consistent with the semi-analytic \texttt{KerrBAM} distributions to $\approx1\sigma$. }
    \label{fig:inc17_grmhd}
\end{figure*}

Finally, we compare the spin and inclination values estimated from the measured interferometric observables to the spread expected from \texttt{KerrBAM}.  For each spin and inclination combination, we compute the average value and spread of the $w_b$ and $d_+$ parameters, which become bands of valid estimates in the $(a, \theta_{\mathrm{o}})$ parameter space. We then compare these spreads to the estimates of spin and inclination performed on the GRMHD simulations. The results are shown for $\theta_{\mathrm{o}} = 30^\circ$ and $\theta_{\mathrm{o}} = 10^\circ$ in \autoref{fig:inc30_grmhd} and \autoref{fig:inc17_grmhd}, respectively. 

We find that the \texttt{KerrBAM}-based framework is able to constrain the GRMHD parameters to $\pm20\%$ in spin and $\pm20^\circ$ in inclination for most of the simulations. For low inclination ($\theta_{\rm o}=10^\circ$), the recovery of inclination is excellent, consistent up to $\approx\!1\sigma$ with the \texttt{KerrBAM} spread. Spin is well-constrained at $a=0.25$ and $a=0.93$ but more poorly constrained at moderate spins ($a=0.5$). We observe a ``pile-up'' effect, particularly around $\theta_{\rm o}\approx5^\circ$, reflecting cases where the posterior distribution of some estimates is constrained by its lower bound. We interpret this behavior to indicate the GRMHD $w_b$ values extend beyond the range predicted by the \texttt{KerrBAM} model and confirm this is the case for several of the tests.

At $\theta_{\rm o}=30^\circ$, spin recovery is comparable to $\theta_{\rm o}=10^\circ$ but inclination recovery is substantially worse. The GRMHD simulations present a significantly larger spread in the measured $w_b$ parameter than predicted by the \texttt{KerrBAM} grid. We attribute this behavior to the significantly more complex structure present in the simulations at moderate inclinations, leading to greater contamination of the $25\textrm{-}35$ G$\lambda$ baseline region with emission not originating from the $n=1$ subring. In particular, SANE simulations with large $R_{\textrm{high}}$ and small $R_{\textrm{low}}$ produce off-equatorial emission, deviating from a key underlying assumption of the \texttt{KerrBAM} model grid. The performance of the framework on GRMHD highlights the challenge of translating patterns identified on semi-analytic or toy-model simulations to more realistic scenarios. However, we note that spin and inclination were somewhat constrained, as the framework could distinguish between high/low spin, and moderate/low inclination, using only the $d_+$ and $w_b$ parameters. The introduction of additional parameters and a more accurate training grid may improve performance to the level required for a 10\% spin constraint with BHEX.

\section{Discussion and conclusions}
\label{sec:conclusions}

In this Letter, we explore a novel method of measuring black hole spin via horizon-scale images from the upcoming Black Hole Explorer mission. 
Given that the brightness asymmetry in black hole photon rings exhibits clear dependence on spin and inclination, we propose that direct measurement of this asymmetry in Fourier space combined with measurements of spin-sensitive quantities (e.g., the maximum diameter $d_+$ of the photon ring) may provide a path to robust constraints on black hole spin. 

We introduce a natural visibility-amplitude-based asymmetry metric, defined by $(V_{\rm max}+V_{\rm min})/(V_{\rm max}-V_{\rm min})$ evaluated on the $n=1$ subring region of dominance, measured as a function of baseline angle $\varphi$. Investigations using simplified semi-analytic models indicate that this particular measure, while intuitive, is sensitive to astrophysical parameters, limiting its utility for measurement. 

As an alternative, we propose direct integration of the visibility amplitudes on the $n=1$ subring region of dominance, which is less sensitive to the parameters of the fluid dynamics. We define a new quantity $w_b$ constructed by taking the width of the angular profile found by integrating the visibility amplitudes over the region of dominance. Our investigation of this quantity on semi-analytic simulations indicates that $w_b$ correlates strongly with inclination across a range of black hole spins and fluid physics parameters. Preliminary validation against general relativistic magnetohydrodynamic (GRMHD) simulations confirms that $w_b$ and $d_+$ can somewhat constrain spin and inclination to $\approx 20\%$.

With these encouraging preliminary findings, the next step is to characterize the space of systematic uncertainties. Our current analysis assumes primarily equatorial emission, a condition valid for \texttt{KerrBAM} and MAD GRMHD scenarios. Additional physical factors governing the fluid flow may complicate the accurate recovery of spin and inclination using the integrated quantity $w_b$. Understanding how deviations from the equatorial emission assumption influence measurement outcomes is therefore essential. Although these uncertainties present challenges, incorporating supplementary observables (e.g., minimum or average diameter, quantities from the $n=0$ subring, etc.) could mitigate degeneracies, thereby enhancing the robustness and accuracy of spin and inclination measurements.

\acknowledgements
J.R.F. is supported by the U.S. National Science Foundation (NSF) Graduate Research Fellowship Program under grant No. 2139319. A.L.\,was supported by NSF grants PHY-2340457 and AST-2307888.
We thank Charles Gammie and his students for providing the GRMHD simulations used in this paper. We acknowledge financial support from the National Science Foundation (AST-2307887). This publication is funded in part by the Gordon and Betty Moore Foundation, Grant GBMF12987. This work was supported by the Black Hole Initiative, which is funded by grants from the John Templeton Foundation (Grant \#62286) and the Gordon and Betty Moore Foundation (Grant GBMF-8273) - although the opinions expressed in this work are those of the author(s) and do not necessarily reflect the views of these Foundations.

\bibliography{SpinInference.bib}

\begin{thebibliography}{25}%
\makeatletter
\providecommand \@ifxundefined [1]{%
 \@ifx{#1\undefined}
}%
\providecommand \@ifnum [1]{%
 \ifnum #1\expandafter \@firstoftwo
 \else \expandafter \@secondoftwo
 \fi
}%
\providecommand \@ifx [1]{%
 \ifx #1\expandafter \@firstoftwo
 \else \expandafter \@secondoftwo
 \fi
}%
\providecommand \natexlab [1]{#1}%
\providecommand \enquote  [1]{``#1''}%
\providecommand \bibnamefont  [1]{#1}%
\providecommand \bibfnamefont [1]{#1}%
\providecommand \citenamefont [1]{#1}%
\providecommand \href@noop [0]{\@secondoftwo}%
\providecommand \href [0]{\begingroup \@sanitize@url \@href}%
\providecommand \@href[1]{\@@startlink{#1}\@@href}%
\providecommand \@@href[1]{\endgroup#1\@@endlink}%
\providecommand \@sanitize@url [0]{\catcode `\\12\catcode `\$12\catcode `\&12\catcode `\#12\catcode `\^12\catcode `\_12\catcode `\%12\relax}%
\providecommand \@@startlink[1]{}%
\providecommand \@@endlink[0]{}%
\providecommand \url  [0]{\begingroup\@sanitize@url \@url }%
\providecommand \@url [1]{\endgroup\@href {#1}{\urlprefix }}%
\providecommand \urlprefix  [0]{URL }%
\providecommand \Eprint [0]{\href }%
\providecommand \doibase [0]{http://dx.doi.org/}%
\providecommand \selectlanguage [0]{\@gobble}%
\providecommand \bibinfo  [0]{\@secondoftwo}%
\providecommand \bibfield  [0]{\@secondoftwo}%
\providecommand \translation [1]{[#1]}%
\providecommand \BibitemOpen [0]{}%
\providecommand \bibitemStop [0]{}%
\providecommand \bibitemNoStop [0]{.\EOS\space}%
\providecommand \EOS [0]{\spacefactor3000\relax}%
\providecommand \BibitemShut  [1]{\csname bibitem#1\endcsname}%
\let\auto@bib@innerbib\@empty
\bibitem [{\citenamefont {{Event Horizon Telescope Collaboration}}(2019)}]{M87I}%
  \BibitemOpen
  \bibfield  {author} {\bibinfo {author} {\bibnamefont {{Event Horizon Telescope Collaboration}}},\ }\href {\doibase 10.3847/2041-8213/ab0ec7} {\bibfield  {journal} {\bibinfo  {journal} {\apjl}\ }\textbf {\bibinfo {volume} {875}},\ \bibinfo {eid} {L1} (\bibinfo {year} {2019})},\ \Eprint {http://arxiv.org/abs/1906.11238} {arXiv:1906.11238 [astro-ph.GA]} \BibitemShut {NoStop}%
\bibitem [{\citenamefont {{Event Horizon Telescope Collaboration}}(2022)}]{SgrAI}%
  \BibitemOpen
  \bibfield  {author} {\bibinfo {author} {\bibnamefont {{Event Horizon Telescope Collaboration}}},\ }\href {\doibase 10.3847/2041-8213/ac6674} {\bibfield  {journal} {\bibinfo  {journal} {\apjl}\ }\textbf {\bibinfo {volume} {930}},\ \bibinfo {eid} {L12} (\bibinfo {year} {2022})}\BibitemShut {NoStop}%
\bibitem [{\citenamefont {{Johnson}}\ \emph {et~al.}(2024)\citenamefont {{Johnson}}, \citenamefont {{Akiyama}}, \citenamefont {{Baturin}}, \citenamefont {{Bilyeu}}, \citenamefont {{Blackburn}}, \citenamefont {{Boroson}}, \citenamefont {{C{\'a}rdenas-Avenda{\~n}o}}, \citenamefont {{Chael}}, \citenamefont {{Chan}}, \citenamefont {{Chang}}, \citenamefont {{Cheimets}}, \citenamefont {{Chou}}, \citenamefont {{Doeleman}}, \citenamefont {{Farah}}, \citenamefont {{Galison}}, \citenamefont {{Gamble}}, \citenamefont {{Gammie}}, \citenamefont {{Gelles}}, \citenamefont {{G{\'o}mez}}, \citenamefont {{Gralla}}, \citenamefont {{Grimes}}, \citenamefont {{Gurvits}}, \citenamefont {{Hadar}}, \citenamefont {{Haworth}}, \citenamefont {{Hada}}, \citenamefont {{Hecht}}, \citenamefont {{Honma}}, \citenamefont {{Houston}}, \citenamefont {{Hudson}}, \citenamefont {{Issaoun}}, \citenamefont {{Jia}}, \citenamefont {{Jorstad}}, \citenamefont {{Kauffman}}, \citenamefont {{Kovalev}}, \citenamefont {{Kurczynski}}, \citenamefont {{Lafon}},
  \citenamefont {{Lupsasca}}, \citenamefont {{Lehmensiek}}, \citenamefont {{Ma}}, \citenamefont {{Marrone}}, \citenamefont {{Marscher}}, \citenamefont {{Melnick}}, \citenamefont {{Narayan}}, \citenamefont {{Niinuma}}, \citenamefont {{Noble}}, \citenamefont {{Palmer}}, \citenamefont {{Palumbo}}, \citenamefont {{Paritsky}}, \citenamefont {{Peretz}}, \citenamefont {{Pesce}}, \citenamefont {{Plavin}}, \citenamefont {{Quataert}}, \citenamefont {{Rana}}, \citenamefont {{Ricarte}}, \citenamefont {{Roelofs}}, \citenamefont {{Shtyrkova}}, \citenamefont {{Sinclair}}, \citenamefont {{Small}}, \citenamefont {{Kumara}}, \citenamefont {{Srinivasan}}, \citenamefont {{Strominger}}, \citenamefont {{Tiede}}, \citenamefont {{Tong}}, \citenamefont {{Wang}}, \citenamefont {{Weintroub}}, \citenamefont {{Wielgus}},\ and\ \citenamefont {{Wong}}}]{BHEXMotivation}%
  \BibitemOpen
  \bibfield  {author} {\bibinfo {author} {\bibfnamefont {M.~D.}\ \bibnamefont {{Johnson}}}, \bibinfo {author} {\bibfnamefont {K.}~\bibnamefont {{Akiyama}}}, \bibinfo {author} {\bibfnamefont {R.}~\bibnamefont {{Baturin}}}, \bibinfo {author} {\bibfnamefont {B.}~\bibnamefont {{Bilyeu}}}, \bibinfo {author} {\bibfnamefont {L.}~\bibnamefont {{Blackburn}}}, \bibinfo {author} {\bibfnamefont {D.}~\bibnamefont {{Boroson}}}, \bibinfo {author} {\bibfnamefont {A.}~\bibnamefont {{C{\'a}rdenas-Avenda{\~n}o}}}, \bibinfo {author} {\bibfnamefont {A.}~\bibnamefont {{Chael}}}, \bibinfo {author} {\bibfnamefont {C.-k.}\ \bibnamefont {{Chan}}}, \bibinfo {author} {\bibfnamefont {D.}~\bibnamefont {{Chang}}}, \bibinfo {author} {\bibfnamefont {P.}~\bibnamefont {{Cheimets}}}, \bibinfo {author} {\bibfnamefont {C.}~\bibnamefont {{Chou}}}, \bibinfo {author} {\bibfnamefont {S.~S.}\ \bibnamefont {{Doeleman}}}, \bibinfo {author} {\bibfnamefont {J.}~\bibnamefont {{Farah}}}, \bibinfo {author} {\bibfnamefont {P.}~\bibnamefont {{Galison}}},
  \bibinfo {author} {\bibfnamefont {R.}~\bibnamefont {{Gamble}}}, \bibinfo {author} {\bibfnamefont {C.~F.}\ \bibnamefont {{Gammie}}}, \bibinfo {author} {\bibfnamefont {Z.}~\bibnamefont {{Gelles}}}, \bibinfo {author} {\bibfnamefont {J.~L.}\ \bibnamefont {{G{\'o}mez}}}, \bibinfo {author} {\bibfnamefont {S.~E.}\ \bibnamefont {{Gralla}}}, \bibinfo {author} {\bibfnamefont {P.}~\bibnamefont {{Grimes}}}, \bibinfo {author} {\bibfnamefont {L.~I.}\ \bibnamefont {{Gurvits}}}, \bibinfo {author} {\bibfnamefont {S.}~\bibnamefont {{Hadar}}}, \bibinfo {author} {\bibfnamefont {K.}~\bibnamefont {{Haworth}}}, \bibinfo {author} {\bibfnamefont {K.}~\bibnamefont {{Hada}}}, \bibinfo {author} {\bibfnamefont {M.~H.}\ \bibnamefont {{Hecht}}}, \bibinfo {author} {\bibfnamefont {M.}~\bibnamefont {{Honma}}}, \bibinfo {author} {\bibfnamefont {J.}~\bibnamefont {{Houston}}}, \bibinfo {author} {\bibfnamefont {B.}~\bibnamefont {{Hudson}}}, \bibinfo {author} {\bibfnamefont {S.}~\bibnamefont {{Issaoun}}}, \bibinfo {author} {\bibfnamefont
  {H.}~\bibnamefont {{Jia}}}, \bibinfo {author} {\bibfnamefont {S.}~\bibnamefont {{Jorstad}}}, \bibinfo {author} {\bibfnamefont {J.}~\bibnamefont {{Kauffman}}}, \bibinfo {author} {\bibfnamefont {Y.~Y.}\ \bibnamefont {{Kovalev}}}, \bibinfo {author} {\bibfnamefont {P.}~\bibnamefont {{Kurczynski}}}, \bibinfo {author} {\bibfnamefont {R.~E.}\ \bibnamefont {{Lafon}}}, \bibinfo {author} {\bibfnamefont {A.}~\bibnamefont {{Lupsasca}}}, \bibinfo {author} {\bibfnamefont {R.}~\bibnamefont {{Lehmensiek}}}, \bibinfo {author} {\bibfnamefont {C.-P.}\ \bibnamefont {{Ma}}}, \bibinfo {author} {\bibfnamefont {D.~P.}\ \bibnamefont {{Marrone}}}, \bibinfo {author} {\bibfnamefont {A.~P.}\ \bibnamefont {{Marscher}}}, \bibinfo {author} {\bibfnamefont {G.}~\bibnamefont {{Melnick}}}, \bibinfo {author} {\bibfnamefont {R.}~\bibnamefont {{Narayan}}}, \bibinfo {author} {\bibfnamefont {K.}~\bibnamefont {{Niinuma}}}, \bibinfo {author} {\bibfnamefont {S.~C.}\ \bibnamefont {{Noble}}}, \bibinfo {author} {\bibfnamefont {E.~J.}\ \bibnamefont
  {{Palmer}}}, \bibinfo {author} {\bibfnamefont {D.~C.~M.}\ \bibnamefont {{Palumbo}}}, \bibinfo {author} {\bibfnamefont {L.}~\bibnamefont {{Paritsky}}}, \bibinfo {author} {\bibfnamefont {E.}~\bibnamefont {{Peretz}}}, \bibinfo {author} {\bibfnamefont {D.}~\bibnamefont {{Pesce}}}, \bibinfo {author} {\bibfnamefont {A.}~\bibnamefont {{Plavin}}}, \bibinfo {author} {\bibfnamefont {E.}~\bibnamefont {{Quataert}}}, \bibinfo {author} {\bibfnamefont {H.}~\bibnamefont {{Rana}}}, \bibinfo {author} {\bibfnamefont {A.}~\bibnamefont {{Ricarte}}}, \bibinfo {author} {\bibfnamefont {F.}~\bibnamefont {{Roelofs}}}, \bibinfo {author} {\bibfnamefont {K.}~\bibnamefont {{Shtyrkova}}}, \bibinfo {author} {\bibfnamefont {L.~C.}\ \bibnamefont {{Sinclair}}}, \bibinfo {author} {\bibfnamefont {J.}~\bibnamefont {{Small}}}, \bibinfo {author} {\bibfnamefont {S.~T.}\ \bibnamefont {{Kumara}}}, \bibinfo {author} {\bibfnamefont {R.}~\bibnamefont {{Srinivasan}}}, \bibinfo {author} {\bibfnamefont {A.}~\bibnamefont {{Strominger}}}, \bibinfo {author}
  {\bibfnamefont {P.}~\bibnamefont {{Tiede}}}, \bibinfo {author} {\bibfnamefont {E.}~\bibnamefont {{Tong}}}, \bibinfo {author} {\bibfnamefont {J.}~\bibnamefont {{Wang}}}, \bibinfo {author} {\bibfnamefont {J.}~\bibnamefont {{Weintroub}}}, \bibinfo {author} {\bibfnamefont {M.}~\bibnamefont {{Wielgus}}}, \ and\ \bibinfo {author} {\bibfnamefont {G.}~\bibnamefont {{Wong}}},\ }in\ \href {\doibase 10.1117/12.3019835} {\emph {\bibinfo {booktitle} {Space Telescopes and Instrumentation 2024: Optical, Infrared, and Millimeter Wave}}},\ \bibinfo {series} {Society of Photo-Optical Instrumentation Engineers (SPIE) Conference Series}, Vol.\ \bibinfo {volume} {13092},\ \bibinfo {editor} {edited by\ \bibinfo {editor} {\bibfnamefont {L.~E.}\ \bibnamefont {{Coyle}}}, \bibinfo {editor} {\bibfnamefont {S.}~\bibnamefont {{Matsuura}}}, \ and\ \bibinfo {editor} {\bibfnamefont {M.~D.}\ \bibnamefont {{Perrin}}}}\ (\bibinfo {year} {2024})\ p.\ \bibinfo {pages} {130922D},\ \Eprint {http://arxiv.org/abs/2406.12917} {arXiv:2406.12917
  [astro-ph.IM]} \BibitemShut {NoStop}%
\bibitem [{\citenamefont {{Marrone}}\ \emph {et~al.}(2024)\citenamefont {{Marrone}}, \citenamefont {{Houston}}, \citenamefont {{Akiyama}}, \citenamefont {{Bilyeu}}, \citenamefont {{Boroson}}, \citenamefont {{Grimes}}, \citenamefont {{Haworth}}, \citenamefont {{Lehmensiek}}, \citenamefont {{Peretz}}, \citenamefont {{Rana}}, \citenamefont {{Sinclair}}, \citenamefont {{Kumara}}, \citenamefont {{Srinivasan}}, \citenamefont {{Tong}}, \citenamefont {{Wang}}, \citenamefont {{Weintroub}},\ and\ \citenamefont {{Johnson}}}]{BHEXInstrument}%
  \BibitemOpen
  \bibfield  {author} {\bibinfo {author} {\bibfnamefont {D.~P.}\ \bibnamefont {{Marrone}}}, \bibinfo {author} {\bibfnamefont {J.}~\bibnamefont {{Houston}}}, \bibinfo {author} {\bibfnamefont {K.}~\bibnamefont {{Akiyama}}}, \bibinfo {author} {\bibfnamefont {B.}~\bibnamefont {{Bilyeu}}}, \bibinfo {author} {\bibfnamefont {D.}~\bibnamefont {{Boroson}}}, \bibinfo {author} {\bibfnamefont {P.}~\bibnamefont {{Grimes}}}, \bibinfo {author} {\bibfnamefont {K.}~\bibnamefont {{Haworth}}}, \bibinfo {author} {\bibfnamefont {R.}~\bibnamefont {{Lehmensiek}}}, \bibinfo {author} {\bibfnamefont {E.}~\bibnamefont {{Peretz}}}, \bibinfo {author} {\bibfnamefont {H.}~\bibnamefont {{Rana}}}, \bibinfo {author} {\bibfnamefont {L.~C.}\ \bibnamefont {{Sinclair}}}, \bibinfo {author} {\bibfnamefont {S.~T.}\ \bibnamefont {{Kumara}}}, \bibinfo {author} {\bibfnamefont {R.}~\bibnamefont {{Srinivasan}}}, \bibinfo {author} {\bibfnamefont {E.}~\bibnamefont {{Tong}}}, \bibinfo {author} {\bibfnamefont {J.}~\bibnamefont {{Wang}}}, \bibinfo {author}
  {\bibfnamefont {J.}~\bibnamefont {{Weintroub}}}, \ and\ \bibinfo {author} {\bibfnamefont {M.~D.}\ \bibnamefont {{Johnson}}},\ }in\ \href {\doibase 10.1117/12.3019589} {\emph {\bibinfo {booktitle} {Space Telescopes and Instrumentation 2024: Optical, Infrared, and Millimeter Wave}}},\ \bibinfo {series} {Society of Photo-Optical Instrumentation Engineers (SPIE) Conference Series}, Vol.\ \bibinfo {volume} {13092},\ \bibinfo {editor} {edited by\ \bibinfo {editor} {\bibfnamefont {L.~E.}\ \bibnamefont {{Coyle}}}, \bibinfo {editor} {\bibfnamefont {S.}~\bibnamefont {{Matsuura}}}, \ and\ \bibinfo {editor} {\bibfnamefont {M.~D.}\ \bibnamefont {{Perrin}}}}\ (\bibinfo {year} {2024})\ p.\ \bibinfo {pages} {130922G},\ \Eprint {http://arxiv.org/abs/2406.10143} {arXiv:2406.10143 [astro-ph.IM]} \BibitemShut {NoStop}%
\bibitem [{\citenamefont {{Lupsasca}}\ \emph {et~al.}(2024)\citenamefont {{Lupsasca}}, \citenamefont {{C{\'a}rdenas-Avenda{\~n}o}}, \citenamefont {{Palumbo}}, \citenamefont {{Johnson}}, \citenamefont {{Gralla}}, \citenamefont {{Marrone}}, \citenamefont {{Galison}}, \citenamefont {{Tiede}},\ and\ \citenamefont {{Keeble}}}]{BHEXScience}%
  \BibitemOpen
  \bibfield  {author} {\bibinfo {author} {\bibfnamefont {A.}~\bibnamefont {{Lupsasca}}}, \bibinfo {author} {\bibfnamefont {A.}~\bibnamefont {{C{\'a}rdenas-Avenda{\~n}o}}}, \bibinfo {author} {\bibfnamefont {D.~C.~M.}\ \bibnamefont {{Palumbo}}}, \bibinfo {author} {\bibfnamefont {M.~D.}\ \bibnamefont {{Johnson}}}, \bibinfo {author} {\bibfnamefont {S.~E.}\ \bibnamefont {{Gralla}}}, \bibinfo {author} {\bibfnamefont {D.~P.}\ \bibnamefont {{Marrone}}}, \bibinfo {author} {\bibfnamefont {P.}~\bibnamefont {{Galison}}}, \bibinfo {author} {\bibfnamefont {P.}~\bibnamefont {{Tiede}}}, \ and\ \bibinfo {author} {\bibfnamefont {L.}~\bibnamefont {{Keeble}}},\ }in\ \href {\doibase 10.1117/12.3019437} {\emph {\bibinfo {booktitle} {Space Telescopes and Instrumentation 2024: Optical, Infrared, and Millimeter Wave}}},\ \bibinfo {series} {Society of Photo-Optical Instrumentation Engineers (SPIE) Conference Series}, Vol.\ \bibinfo {volume} {13092},\ \bibinfo {editor} {edited by\ \bibinfo {editor} {\bibfnamefont {L.~E.}\ \bibnamefont
  {{Coyle}}}, \bibinfo {editor} {\bibfnamefont {S.}~\bibnamefont {{Matsuura}}}, \ and\ \bibinfo {editor} {\bibfnamefont {M.~D.}\ \bibnamefont {{Perrin}}}}\ (\bibinfo {year} {2024})\ p.\ \bibinfo {pages} {130926Q}\BibitemShut {NoStop}%
\bibitem [{\citenamefont {Lockhart}\ and\ \citenamefont {Gralla}(2022)}]{Lockhart_2022}%
  \BibitemOpen
  \bibfield  {author} {\bibinfo {author} {\bibfnamefont {W.}~\bibnamefont {Lockhart}}\ and\ \bibinfo {author} {\bibfnamefont {S.~E.}\ \bibnamefont {Gralla}},\ }\href {\doibase 10.1093/mnras/stac2743} {\bibfield  {journal} {\bibinfo  {journal} {Monthly Notices of the Royal Astronomical Society}\ }\textbf {\bibinfo {volume} {517}},\ \bibinfo {pages} {2462–2470} (\bibinfo {year} {2022})}\BibitemShut {NoStop}%
\bibitem [{\citenamefont {Tiede}\ \emph {et~al.}(2022)\citenamefont {Tiede}, \citenamefont {Johnson}, \citenamefont {Pesce}, \citenamefont {Palumbo}, \citenamefont {Chang},\ and\ \citenamefont {Galison}}]{Tiede2022}%
  \BibitemOpen
  \bibfield  {author} {\bibinfo {author} {\bibfnamefont {P.}~\bibnamefont {Tiede}}, \bibinfo {author} {\bibfnamefont {M.~D.}\ \bibnamefont {Johnson}}, \bibinfo {author} {\bibfnamefont {D.~W.}\ \bibnamefont {Pesce}}, \bibinfo {author} {\bibfnamefont {D.~C.~M.}\ \bibnamefont {Palumbo}}, \bibinfo {author} {\bibfnamefont {D.~O.}\ \bibnamefont {Chang}}, \ and\ \bibinfo {author} {\bibfnamefont {P.}~\bibnamefont {Galison}},\ }\href {https://arxiv.org/abs/2210.13498} {\enquote {\bibinfo {title} {Measuring photon rings with the ngeht},}\ } (\bibinfo {year} {2022}),\ \Eprint {http://arxiv.org/abs/2210.13498} {arXiv:2210.13498 [astro-ph.HE]} \BibitemShut {NoStop}%
\bibitem [{\citenamefont {{Luminet}}(1979)}]{Luminet1979}%
  \BibitemOpen
  \bibfield  {author} {\bibinfo {author} {\bibfnamefont {J.~P.}\ \bibnamefont {{Luminet}}},\ }\href@noop {} {\bibfield  {journal} {\bibinfo  {journal} {\aap}\ }\textbf {\bibinfo {volume} {75}},\ \bibinfo {pages} {228} (\bibinfo {year} {1979})}\BibitemShut {NoStop}%
\bibitem [{\citenamefont {{Gralla}}\ \emph {et~al.}(2019)\citenamefont {{Gralla}}, \citenamefont {{Holz}},\ and\ \citenamefont {{Wald}}}]{GrallaHolzWald}%
  \BibitemOpen
  \bibfield  {author} {\bibinfo {author} {\bibfnamefont {S.~E.}\ \bibnamefont {{Gralla}}}, \bibinfo {author} {\bibfnamefont {D.~E.}\ \bibnamefont {{Holz}}}, \ and\ \bibinfo {author} {\bibfnamefont {R.~M.}\ \bibnamefont {{Wald}}},\ }\href {\doibase 10.1103/PhysRevD.100.024018} {\bibfield  {journal} {\bibinfo  {journal} {\prd}\ }\textbf {\bibinfo {volume} {100}},\ \bibinfo {eid} {024018} (\bibinfo {year} {2019})},\ \Eprint {http://arxiv.org/abs/1906.00873} {arXiv:1906.00873 [astro-ph.HE]} \BibitemShut {NoStop}%
\bibitem [{\citenamefont {{Johnson}}\ \emph {et~al.}(2020)\citenamefont {{Johnson}}, \citenamefont {{Lupsasca}}, \citenamefont {{Strominger}}, \citenamefont {{Wong}}, \citenamefont {{Hadar}}, \citenamefont {{Kapec}}, \citenamefont {{Narayan}}, \citenamefont {{Chael}}, \citenamefont {{Gammie}}, \citenamefont {{Galison}}, \citenamefont {{Palumbo}}, \citenamefont {{Doeleman}}, \citenamefont {{Blackburn}}, \citenamefont {{Wielgus}}, \citenamefont {{Pesce}}, \citenamefont {{Farah}},\ and\ \citenamefont {{Moran}}}]{JohnsonLupsasca2020}%
  \BibitemOpen
  \bibfield  {author} {\bibinfo {author} {\bibfnamefont {M.~D.}\ \bibnamefont {{Johnson}}}, \bibinfo {author} {\bibfnamefont {A.}~\bibnamefont {{Lupsasca}}}, \bibinfo {author} {\bibfnamefont {A.}~\bibnamefont {{Strominger}}}, \bibinfo {author} {\bibfnamefont {G.~N.}\ \bibnamefont {{Wong}}}, \bibinfo {author} {\bibfnamefont {S.}~\bibnamefont {{Hadar}}}, \bibinfo {author} {\bibfnamefont {D.}~\bibnamefont {{Kapec}}}, \bibinfo {author} {\bibfnamefont {R.}~\bibnamefont {{Narayan}}}, \bibinfo {author} {\bibfnamefont {A.}~\bibnamefont {{Chael}}}, \bibinfo {author} {\bibfnamefont {C.~F.}\ \bibnamefont {{Gammie}}}, \bibinfo {author} {\bibfnamefont {P.}~\bibnamefont {{Galison}}}, \bibinfo {author} {\bibfnamefont {D.~C.~M.}\ \bibnamefont {{Palumbo}}}, \bibinfo {author} {\bibfnamefont {S.~S.}\ \bibnamefont {{Doeleman}}}, \bibinfo {author} {\bibfnamefont {L.}~\bibnamefont {{Blackburn}}}, \bibinfo {author} {\bibfnamefont {M.}~\bibnamefont {{Wielgus}}}, \bibinfo {author} {\bibfnamefont {D.~W.}\ \bibnamefont {{Pesce}}},
  \bibinfo {author} {\bibfnamefont {J.~R.}\ \bibnamefont {{Farah}}}, \ and\ \bibinfo {author} {\bibfnamefont {J.~M.}\ \bibnamefont {{Moran}}},\ }\href {\doibase 10.1126/sciadv.aaz1310} {\bibfield  {journal} {\bibinfo  {journal} {Science Advances}\ }\textbf {\bibinfo {volume} {6}},\ \bibinfo {pages} {eaaz1310} (\bibinfo {year} {2020})},\ \Eprint {http://arxiv.org/abs/1907.04329} {arXiv:1907.04329 [astro-ph.IM]} \BibitemShut {NoStop}%
\bibitem [{\citenamefont {{Gralla}}\ and\ \citenamefont {{Lupsasca}}(2020)}]{GrallaLupsascaLensing}%
  \BibitemOpen
  \bibfield  {author} {\bibinfo {author} {\bibfnamefont {S.~E.}\ \bibnamefont {{Gralla}}}\ and\ \bibinfo {author} {\bibfnamefont {A.}~\bibnamefont {{Lupsasca}}},\ }\href {\doibase 10.1103/PhysRevD.101.044031} {\bibfield  {journal} {\bibinfo  {journal} {\prd}\ }\textbf {\bibinfo {volume} {101}},\ \bibinfo {eid} {044031} (\bibinfo {year} {2020})},\ \Eprint {http://arxiv.org/abs/1910.12873} {arXiv:1910.12873 [gr-qc]} \BibitemShut {NoStop}%
\bibitem [{\citenamefont {{C{\'a}rdenas-Avenda{\~n}o}}\ and\ \citenamefont {{Lupsasca}}(2023)}]{Cardenas2023}%
  \BibitemOpen
  \bibfield  {author} {\bibinfo {author} {\bibfnamefont {A.}~\bibnamefont {{C{\'a}rdenas-Avenda{\~n}o}}}\ and\ \bibinfo {author} {\bibfnamefont {A.}~\bibnamefont {{Lupsasca}}},\ }\href {\doibase 10.1103/PhysRevD.108.064043} {\bibfield  {journal} {\bibinfo  {journal} {\prd}\ }\textbf {\bibinfo {volume} {108}},\ \bibinfo {eid} {064043} (\bibinfo {year} {2023})},\ \Eprint {http://arxiv.org/abs/2305.12956} {arXiv:2305.12956 [gr-qc]} \BibitemShut {NoStop}%
\bibitem [{\citenamefont {{Gammie}}\ \emph {et~al.}(2003)\citenamefont {{Gammie}}, \citenamefont {{McKinney}},\ and\ \citenamefont {{T{\'o}th}}}]{Gammie_2003}%
  \BibitemOpen
  \bibfield  {author} {\bibinfo {author} {\bibfnamefont {C.~F.}\ \bibnamefont {{Gammie}}}, \bibinfo {author} {\bibfnamefont {J.~C.}\ \bibnamefont {{McKinney}}}, \ and\ \bibinfo {author} {\bibfnamefont {G.}~\bibnamefont {{T{\'o}th}}},\ }\href {\doibase 10.1086/374594} {\bibfield  {journal} {\bibinfo  {journal} {\apj}\ }\textbf {\bibinfo {volume} {589}},\ \bibinfo {pages} {444} (\bibinfo {year} {2003})},\ \Eprint {http://arxiv.org/abs/astro-ph/0301509} {arXiv:astro-ph/0301509 [astro-ph]} \BibitemShut {NoStop}%
\bibitem [{\citenamefont {{Chang}}\ \emph {et~al.}(2024)\citenamefont {{Chang}}, \citenamefont {{Johnson}}, \citenamefont {{Tiede}},\ and\ \citenamefont {{Palumbo}}}]{BlackHolePhotogrammetry}%
  \BibitemOpen
  \bibfield  {author} {\bibinfo {author} {\bibfnamefont {D.~O.}\ \bibnamefont {{Chang}}}, \bibinfo {author} {\bibfnamefont {M.~D.}\ \bibnamefont {{Johnson}}}, \bibinfo {author} {\bibfnamefont {P.}~\bibnamefont {{Tiede}}}, \ and\ \bibinfo {author} {\bibfnamefont {D.~C.~M.}\ \bibnamefont {{Palumbo}}},\ }\href {\doibase 10.3847/1538-4357/ad6b28} {\bibfield  {journal} {\bibinfo  {journal} {\apj}\ }\textbf {\bibinfo {volume} {974}},\ \bibinfo {eid} {143} (\bibinfo {year} {2024})}\BibitemShut {NoStop}%
\bibitem [{\citenamefont {{Palumbo}}\ \emph {et~al.}(2022)\citenamefont {{Palumbo}}, \citenamefont {{Gelles}}, \citenamefont {{Tiede}}, \citenamefont {{Chang}}, \citenamefont {{Pesce}}, \citenamefont {{Chael}},\ and\ \citenamefont {{Johnson}}}]{KerrBAM}%
  \BibitemOpen
  \bibfield  {author} {\bibinfo {author} {\bibfnamefont {D.~C.~M.}\ \bibnamefont {{Palumbo}}}, \bibinfo {author} {\bibfnamefont {Z.}~\bibnamefont {{Gelles}}}, \bibinfo {author} {\bibfnamefont {P.}~\bibnamefont {{Tiede}}}, \bibinfo {author} {\bibfnamefont {D.~O.}\ \bibnamefont {{Chang}}}, \bibinfo {author} {\bibfnamefont {D.~W.}\ \bibnamefont {{Pesce}}}, \bibinfo {author} {\bibfnamefont {A.}~\bibnamefont {{Chael}}}, \ and\ \bibinfo {author} {\bibfnamefont {M.~D.}\ \bibnamefont {{Johnson}}},\ }\href {\doibase 10.3847/1538-4357/ac9ab7} {\bibfield  {journal} {\bibinfo  {journal} {\apj}\ }\textbf {\bibinfo {volume} {939}},\ \bibinfo {eid} {107} (\bibinfo {year} {2022})},\ \Eprint {http://arxiv.org/abs/2210.07108} {arXiv:2210.07108 [astro-ph.HE]} \BibitemShut {NoStop}%
\bibitem [{\citenamefont {{Broderick}}\ \emph {et~al.}(2022)\citenamefont {{Broderick}}, \citenamefont {{Tiede}}, \citenamefont {{Pesce}},\ and\ \citenamefont {{Gold}}}]{Broderick2022}%
  \BibitemOpen
  \bibfield  {author} {\bibinfo {author} {\bibfnamefont {A.~E.}\ \bibnamefont {{Broderick}}}, \bibinfo {author} {\bibfnamefont {P.}~\bibnamefont {{Tiede}}}, \bibinfo {author} {\bibfnamefont {D.~W.}\ \bibnamefont {{Pesce}}}, \ and\ \bibinfo {author} {\bibfnamefont {R.}~\bibnamefont {{Gold}}},\ }\href {\doibase 10.3847/1538-4357/ac4970} {\bibfield  {journal} {\bibinfo  {journal} {\apj}\ }\textbf {\bibinfo {volume} {927}},\ \bibinfo {eid} {6} (\bibinfo {year} {2022})},\ \Eprint {http://arxiv.org/abs/2105.09962} {arXiv:2105.09962 [astro-ph.HE]} \BibitemShut {NoStop}%
\bibitem [{\citenamefont {Farah}\ \emph {et~al.}(2024)\citenamefont {Farah}, \citenamefont {Davelaar}, \citenamefont {Palumbo}, \citenamefont {Johnson},\ and\ \citenamefont {Delgado}}]{Farah:PR1}%
  \BibitemOpen
  \bibfield  {author} {\bibinfo {author} {\bibfnamefont {J.~R.}\ \bibnamefont {Farah}}, \bibinfo {author} {\bibfnamefont {J.}~\bibnamefont {Davelaar}}, \bibinfo {author} {\bibfnamefont {D.}~\bibnamefont {Palumbo}}, \bibinfo {author} {\bibfnamefont {M.~D.}\ \bibnamefont {Johnson}}, \ and\ \bibinfo {author} {\bibfnamefont {J.}~\bibnamefont {Delgado}},\ }\href {https://arxiv.org/abs/2411.01060} {\enquote {\bibinfo {title} {Machine- and deep-learning-driven angular momentum inference from bhex observations of the $n=1$ photon ring},}\ } (\bibinfo {year} {2024}),\ \Eprint {http://arxiv.org/abs/2411.01060} {arXiv:2411.01060 [astro-ph.HE]} \BibitemShut {NoStop}%
\bibitem [{\citenamefont {{Palumbo}}\ and\ \citenamefont {{Wong}}(2022)}]{LinearPolarization}%
  \BibitemOpen
  \bibfield  {author} {\bibinfo {author} {\bibfnamefont {D.~C.~M.}\ \bibnamefont {{Palumbo}}}\ and\ \bibinfo {author} {\bibfnamefont {G.~N.}\ \bibnamefont {{Wong}}},\ }\href {\doibase 10.3847/1538-4357/ac59b4} {\bibfield  {journal} {\bibinfo  {journal} {\apj}\ }\textbf {\bibinfo {volume} {929}},\ \bibinfo {eid} {49} (\bibinfo {year} {2022})},\ \Eprint {http://arxiv.org/abs/2203.00844} {arXiv:2203.00844 [astro-ph.HE]} \BibitemShut {NoStop}%
\bibitem [{\citenamefont {{Paugnat}}\ \emph {et~al.}(2022)\citenamefont {{Paugnat}}, \citenamefont {{Lupsasca}}, \citenamefont {{Vincent}},\ and\ \citenamefont {{Wielgus}}}]{Paugnat2022}%
  \BibitemOpen
  \bibfield  {author} {\bibinfo {author} {\bibfnamefont {H.}~\bibnamefont {{Paugnat}}}, \bibinfo {author} {\bibfnamefont {A.}~\bibnamefont {{Lupsasca}}}, \bibinfo {author} {\bibfnamefont {F.~H.}\ \bibnamefont {{Vincent}}}, \ and\ \bibinfo {author} {\bibfnamefont {M.}~\bibnamefont {{Wielgus}}},\ }\href {\doibase 10.1051/0004-6361/202244216} {\bibfield  {journal} {\bibinfo  {journal} {\aap}\ }\textbf {\bibinfo {volume} {668}},\ \bibinfo {eid} {A11} (\bibinfo {year} {2022})},\ \Eprint {http://arxiv.org/abs/2206.02781} {arXiv:2206.02781 [astro-ph.HE]} \BibitemShut {NoStop}%
\bibitem [{\citenamefont {{Keeble}}\ \emph {et~al.}(2025)\citenamefont {{Keeble}}, \citenamefont {{C{\'a}rdenas-Avenda{\~n}o}},\ and\ \citenamefont {{Palumbo}}}]{Keeble2025}%
  \BibitemOpen
  \bibfield  {author} {\bibinfo {author} {\bibfnamefont {L.~S.}\ \bibnamefont {{Keeble}}}, \bibinfo {author} {\bibfnamefont {A.}~\bibnamefont {{C{\'a}rdenas-Avenda{\~n}o}}}, \ and\ \bibinfo {author} {\bibfnamefont {D.~C.~M.}\ \bibnamefont {{Palumbo}}},\ }\href {\doibase 10.48550/arXiv.2502.20312} {\bibfield  {journal} {\bibinfo  {journal} {arXiv e-prints}\ ,\ \bibinfo {eid} {arXiv:2502.20312}} (\bibinfo {year} {2025})},\ \Eprint {http://arxiv.org/abs/2502.20312} {arXiv:2502.20312 [astro-ph.HE]} \BibitemShut {NoStop}%
\bibitem [{\citenamefont {{Gralla}}(2020)}]{Gralla2020}%
  \BibitemOpen
  \bibfield  {author} {\bibinfo {author} {\bibfnamefont {S.~E.}\ \bibnamefont {{Gralla}}},\ }\href {\doibase 10.1103/PhysRevD.102.044017} {\bibfield  {journal} {\bibinfo  {journal} {\prd}\ }\textbf {\bibinfo {volume} {102}},\ \bibinfo {eid} {044017} (\bibinfo {year} {2020})},\ \Eprint {http://arxiv.org/abs/2005.03856} {arXiv:2005.03856 [astro-ph.HE]} \BibitemShut {NoStop}%
\bibitem [{\citenamefont {Gralla}\ \emph {et~al.}(2020)\citenamefont {Gralla}, \citenamefont {Lupsasca},\ and\ \citenamefont {Marrone}}]{Gralla:PR1}%
  \BibitemOpen
  \bibfield  {author} {\bibinfo {author} {\bibfnamefont {S.~E.}\ \bibnamefont {Gralla}}, \bibinfo {author} {\bibfnamefont {A.}~\bibnamefont {Lupsasca}}, \ and\ \bibinfo {author} {\bibfnamefont {D.~P.}\ \bibnamefont {Marrone}},\ }\href {\doibase 10.1103/PhysRevD.102.124004} {\bibfield  {journal} {\bibinfo  {journal} {Phys. Rev. D}\ }\textbf {\bibinfo {volume} {102}},\ \bibinfo {pages} {124004} (\bibinfo {year} {2020})},\ \Eprint {http://arxiv.org/abs/2008.03879} {arXiv:2008.03879 [gr-qc]} \BibitemShut {NoStop}%
\bibitem [{\citenamefont {Gralla}\ and\ \citenamefont {Lupsasca}(2020)}]{Gralla:PR2}%
  \BibitemOpen
  \bibfield  {author} {\bibinfo {author} {\bibfnamefont {S.~E.}\ \bibnamefont {Gralla}}\ and\ \bibinfo {author} {\bibfnamefont {A.}~\bibnamefont {Lupsasca}},\ }\href {\doibase 10.1103/PhysRevD.102.124003} {\bibfield  {journal} {\bibinfo  {journal} {Phys. Rev. D}\ }\textbf {\bibinfo {volume} {102}},\ \bibinfo {pages} {124003} (\bibinfo {year} {2020})},\ \Eprint {http://arxiv.org/abs/2007.10336} {arXiv:2007.10336 [gr-qc]} \BibitemShut {NoStop}%
\bibitem [{\citenamefont {{C{\'a}rdenas-Avenda{\~n}o}}\ \emph {et~al.}(2023)\citenamefont {{C{\'a}rdenas-Avenda{\~n}o}}, \citenamefont {{Lupsasca}},\ and\ \citenamefont {{Zhu}}}]{AART}%
  \BibitemOpen
  \bibfield  {author} {\bibinfo {author} {\bibfnamefont {A.}~\bibnamefont {{C{\'a}rdenas-Avenda{\~n}o}}}, \bibinfo {author} {\bibfnamefont {A.}~\bibnamefont {{Lupsasca}}}, \ and\ \bibinfo {author} {\bibfnamefont {H.}~\bibnamefont {{Zhu}}},\ }\href {\doibase 10.1103/PhysRevD.107.043030} {\bibfield  {journal} {\bibinfo  {journal} {\prd}\ }\textbf {\bibinfo {volume} {107}},\ \bibinfo {eid} {043030} (\bibinfo {year} {2023})},\ \Eprint {http://arxiv.org/abs/2211.07469} {arXiv:2211.07469 [gr-qc]} \BibitemShut {NoStop}%
\bibitem [{\citenamefont {{VanderPlas}}(2018)}]{LombScargleReview}%
  \BibitemOpen
  \bibfield  {author} {\bibinfo {author} {\bibfnamefont {J.~T.}\ \bibnamefont {{VanderPlas}}},\ }\href {\doibase 10.3847/1538-4365/aab766} {\bibfield  {journal} {\bibinfo  {journal} {\apjs}\ }\textbf {\bibinfo {volume} {236}},\ \bibinfo {eid} {16} (\bibinfo {year} {2018})},\ \Eprint {http://arxiv.org/abs/1703.09824} {arXiv:1703.09824 [astro-ph.IM]} \BibitemShut {NoStop}%
\end{thebibliography}%

\appendix

\section{Details of observable measurement}
\label{appendix:observable_measurement}

Here, we review the measurement of the $d_+$ and $w_b$ observables in more detail. 

\subsection{Construction of \texorpdfstring{$d_+$}{d+}}

The $d_+$ observable is a measurement of the maximum angular diameter of the $n=1$ subring as a function of azimuthal angle. This observable is directly inferred from the periodicity of oscillations in the visibility function. \cite{Gralla:PR1} and \cite{JohnsonLupsasca2020} both showed that the visibility function on baselines corresponding to length scales much larger than the subring width can be approximated with a sinusoid whose wavelength varies with $1/d_\varphi$. We directly estimate the location of peaks and nulls in the visibility function and compute an average wavelength $\Lambda_V$. The estimated wavelength is then converted to a diameter measurement via
\begin{equation}
    d_\varphi \approx 10\ M \left(\frac{\Lambda_V}{5.7 \text{ G}\lambda}\right)^{-1}, 
\end{equation}
where 5.7 G$\lambda$ corresponds to the $\approx 10\ M$ photon ring diameter at the distance of M87${}^\ast$. This procedure is sensitive to the resolution of the original image. We test for convergence of the measured diameter and find that convergence is achieved at a resolution $>$200x200 pixels, which we use for all simulations generated with \texttt{KerrBAM}. In addition, when numerically calculating the Fourier transform on the measurement region $25 \text{ G}\lambda < u < 35 \text{ G}\lambda$, the $d_+$ quantity is sensitive to the resolution in frequency. To balance performance and accuracy, we sample $N_{\text{res}}=100$ evenly spaced points on the measurement region. This results in a $\approx0.1 \text{ G}\lambda$ systematic error, corresponding to $\approx0.4\%$ of the characteristic diameter. 
To reduce this systematic error even further, we sample several points around the peaks and nulls in the visibility function and compute an expectation value for each peak location, which is then used to compute the wavelength. In our testing, we found that this reduces the resolution systematic error to $\ll0.2\%$ and is equivalent to performing the wavelength estimation with $N_{\text{res}}\gtrsim300$ ($\lesssim0.033 \text{ G}\lambda$ spacing).

The increased complexity of the GRMHD simulations combined with their lower resolution made it more challenging to consistently extract diameters. Given the low number of GRMHD tests in our study, we individually checked the accuracy and performance of each fit and attempted multiple methods of diameter measurement. These methods included (i) using the frequency of a best-fit sinusoid plus linear offset, and (ii) using the peak frequency of a Lomb-Scargle periodogram (see e.g., \cite{LombScargleReview} for a review). Consistent measurement of the photon ring diameter in a complex and noisy astrophysical environment is still an open problem, which we do not attempt to present a robust pipeline for here.

\vspace{0.5cm}
\subsection{Construction of \texorpdfstring{$w_b$}{wb}}

A recent analysis \cite{Farah:PR1} has suggested that the azimuthal brightness distribution of the $n=1$ subring varies with the spin and inclination of the black hole. A convenient proxy to this quantity is the definite integral over the squared visibility amplitudes in the region $u_- < u < u_+$ of $n=1$ domination:
\begin{equation}
    b_\varphi \equiv \int_{u_-}^{u_+} du \ |V(u,\varphi)|^2.
\end{equation}
We compute this quantity for 20 angles $0 < \varphi < \pi$ in the Fourier plane. We find that the distribution of $b_\varphi$ is Gaussian-like with Fourier angle $\varphi$. For the purposes of this analysis, we characterize this distribution based on its FWHM. Similarly to the $d_+$ quantity, we compute $b_\varphi$ and fit the resulting distribution to a Gaussian, which will have an associated FWHM, using which we set $w_b$. In contrast to the diameter measurement, measurement of $w_b$ was of comparable difficulty in both the \texttt{KerrBAM} and GRMHD cases, and the same pipeline was used on both simulation classes.

\subsection{Construction of parameter heatmaps}

For each combination of spin magnitude and inclination, there are several simulations that must be considered. These simulations feature each combination of the fluid parameters, and positive/negative spin. We calculate $d_+$ and $w_b$ for each simulation, creating a distribution for both for each spin/inclination combination. We perform a single outlier-rejection step, rejecting any values which lie $>5\sigma$ away from the average of the bin. We then calculate the median and standard deviation of the remaining values, which form the heatmaps shown in \autoref{fig:dmax} and \autoref{fig:omega_b}. For display purposes, these heatmaps feature a small amount of global Gaussian smoothing and interpolation to remove minor artifacts that are unlikely to correspond to real behavior.

\section{Derivation of \texorpdfstring{$b_\varphi$}{bphi}}

We seek to explore a novel quantity heuristically identified to be a potential inclination discriminant for black hole images. Specifically, recent work has identified the approximately-Gaussian azimuthal brightness profile of the $n=1$ photon ring as varying almost directly with the inclination of the black hole relative to the observer. We seek to find a Fourier representation of this quantity which can be measured in the $n=1$ region of dominance. 

We begin with an image-domain representation of the $n=1$ subring $I(r, \varphi)$. We can calculate a projection of this image function in an arbitrary direction $\varphi$, $\mathcal{P}_\varphi$. This projection can be performed via e.g., the Radon transform.
Once operating on projections, we can exploit the \textit{projection slice theorem}.
We define a new quantity $B_\varphi^{(k)}$ which tracks moments of the total intensity of a projection $\mathcal{P}_{\varphi}$,
\begin{align}
    B_\varphi^{(k)} = \int_{-\infty}^{\infty} ds \ \big[\mathcal{P}_{\varphi}I(s)\big]^k.
\end{align}
We can relate this definition to the total intensity of the image. The zero frequency component of the Fourier transform of $I(r,\varphi)$, denoted $V(0)$, necessarily is the total intensity of the image $I_{\text{tot}}$, as no spatial scale can contain more flux than the entire image. This property is direction independent, since $V(0, \varphi) = V(0)$; i.e., the total intensity of the image $I_{\text{tot}}$ is independent of direction. Therefore, since the projection of an image along any axis will have the same total intensity as the full image,
    \begin{align}
        B_\varphi^{(1)} &=  \int_{-\infty}^{\infty} ds \ \big[\mathcal{P}_{\varphi}I(s)\big] = \int d^2 \vec{x} \ I(\vec{x}) = I_{\text{tot}} = V(0).
    \end{align}
We next make use of \textit{Plancherel's theorem} to investigate a special feature of $B_\varphi^{(k)}$. 
Plancherel's theorem naturally motivates examining the $B_\varphi^{(2)}$ quantity. We consider a slice of the two-dimensional Fourier transform of our image function $I(r, \varphi)$, which we define to be our visibility function:
\begin{equation}
    V(u, \varphi) \equiv \mathcal{F}_{2, r\to u} I(r, \varphi).
\end{equation}
By taking the squared amplitude of this quantity (which can be measured by an interferometer), we can immediately invoke both Plancherel's theorem and the projection slice theorem to state that 
\begin{equation}
    \int_{-\infty}^{\infty} du \ |V(u, \varphi)|^2 = \int_{-\infty}^{\infty} dr \ |\mathcal{P}_\varphi I(r, \varphi)|^2 = B_\varphi^{(2)}.
\end{equation} 
This proves that integrating the Fourier transform over the entire range of frequencies for a particular angle $\varphi$ will yield information about $B_\varphi^{(2)}$ and therefore the azimuthal brightness distribution of the subring. However, an interferometer will not have access to the full range of frequencies. Even more constraining, the $n=1$ subring will only dominate on a range of frequencies $u_- < u < u_+$. Therefore, we must demonstrate that this integral still scales with the total value of $B_\varphi^{(2)}$ even when integrated on a small, finite range.  

For a closed convex curve with diameter $d_\varphi$ (e.g., the $n=1$ subring), we define a new quantity $b_\varphi$, for which
    \begin{equation}
        B_\varphi^{(2)} \propto b_\varphi \equiv \int_{u_-}^{u_+} du \ |V(u, \varphi)|^2, 
    \end{equation}
    where $u_- < u < u_+$ is the $n=1$ subring region of dominance, corresponding to baselines much longer than $1/d_\varphi$.

\noindent \textbf{Proof.} It has been shown that the visibility function of a closed convex curve can be approximated as
\begin{equation}
    |V(u, \varphi)| \approx \frac{1}{\sqrt{u}}\sqrt{(\alpha_\varphi^L)^2 + (\alpha_\varphi^R)^2 + 2\alpha_\varphi^L\alpha_\varphi^R\sin(2\pi d_\varphi u)},
\end{equation}
where $\alpha_\varphi^L$ [$\alpha_\varphi^R$] corresponds to the product of the integrated intensity $I_\varphi$ and square root of the radius of curvature $\sqrt{\mathcal{R}}$ on the left [right] edge of the closed convex curve, and $d_\varphi$ corresponds to the diameter of the curve in the same direction. 

Let us now use this approximation to calculate $b_\varphi$:
\begin{align}
    b_\varphi &= \int_{u_-}^{u_+} du \ |V(u, \varphi)|^2 \nonumber \\
    &= \int_{u_-}^{u_+} du \ \left|\frac{1}{\sqrt{u}}\sqrt{(\alpha_\varphi^L)^2 + (\alpha_\varphi^R)^2 + 2\alpha_\varphi^L\alpha_\varphi^R\sin(2\pi d_\varphi u)}\right|^2 \nonumber \\
    &=  \int_{u_-}^{u_+} du \ \frac{1}{u}\Big[(\alpha_\varphi^L)^2 + (\alpha_\varphi^R)^2 + 2\alpha_\varphi^L\alpha_\varphi^R\sin(2\pi d_\varphi u)\Big] \nonumber \\
    &= \int_{u_-}^{u_+} du \frac{(\alpha_\varphi^L)^2 + (\alpha_\varphi^R)^2}{u} +  \int_{u_-}^{u_+} du \frac{2\alpha_\varphi^L\alpha_\varphi^R\sin(2\pi d_\varphi u)}{u} \nonumber \\
    &= \Big[(\alpha_\varphi^L)^2 + (\alpha_\varphi^R)^2\Big]\log\left(\frac{u_+}{u_-}\right)\nonumber\\
    &\phantom{=}+2\alpha_\varphi^L\alpha_\varphi^R\left[ \text{Si}(2\pi d_\varphi u_+) - \text{Si}(2\pi d_\varphi u_-) \right].
\end{align}
For large values (and we are at very long baselines), the sine integral function $\text{Si}(x)$ tends to a constant value, which allows us to drop the second term in our expression and simply write
\begin{equation}
    b_\varphi \approx \Big[(\alpha_\varphi^L)^2 + (\alpha_\varphi^R)^2\Big]\log\left(\frac{u_+}{u_-}\right).
    \label{eq:bvarphi_prop}
\end{equation}
In the limit of integrating over all frequencies, we have shown (\autoref{eq:bvarphi_vis}) that
\begin{equation}
    B_\varphi^{(2)} = \lim_{u_\pm\to\pm\infty}b_\varphi,
\end{equation}
and since \autoref{eq:bvarphi_prop} shows that $b_\varphi \propto \lim_{\substack{u_+\to \infty \\ u_-\to 0}}b_\varphi$ (scaling precisely by $\log[u_+/u_-]$), we have therefore shown that
\begin{equation}
    B_\varphi^{(2)} \propto b_\varphi.
\end{equation}
Thus, we can make measurements of $b_\varphi$ on the $n=1$ subring region of dominance with BHEX, and in fact probe the azimuthal brightness distribution of the $n=1$ subring, which varies directly with the inclination of the black hole. \vspace{0.25cm}

The construction of $b_\varphi$ can also help us intuitively understand why this quantity seems to directly track the azimuthal brightness distribution of the $n=1$ subring and not simply the more abstract $B_\varphi$ quantity.

Earlier, we showed that 
\begin{equation}
    b_\varphi \propto (\alpha_\varphi^L)^2 + (\alpha_\varphi^R)^2.
\end{equation}
The explicit definition of the $\alpha_\varphi$ quantity is
\begin{equation}
    \alpha_{\varphi}^{L,R} = I_\varphi^{L,R} \sqrt{\mathcal{R}},
\end{equation}
where $\mathcal{R}$ is the radius of curvature of the closed convex curve. However, the $n=1$ subring has a very low ellipticity ($\lesssim 6$\% in the most extreme case), meaning that the radius of curvature is largely constant (certainly significantly less variable than the intensity $I_\varphi$) and approximately simply the radius of the subring, $\mathcal{R}\approx R$ for all $\varphi$. This lets us treat $\mathcal{R}$ as constant, and write
\begin{equation}
    b_\varphi \propto (\alpha_\varphi^L)^2 + (\alpha_\varphi^R)^2 \propto (I_\varphi^{L})^2 + (I_\varphi^{R})^2.
\end{equation}
From this expression, it becomes clear that the quantity $b_\varphi$ is not just analogous but in fact a \textit{direct} probe of the azimuthal brightness distribution.

\end{document}